\documentclass[12pt,letterpaper]{article}\usepackage[]{graphicx}\usepackage[]{color}
\makeatletter
\def\maxwidth{ %
  \ifdim\Gin@nat@width>\linewidth
    \linewidth
  \else
    \Gin@nat@width
  \fi
}
\makeatother

\definecolor{fgcolor}{rgb}{0.345, 0.345, 0.345}

\usepackage{framed}
\makeatletter
 {\par\unskip\endMakeFramed%
 \at@end@of@kframe}
\makeatother

\definecolor{shadecolor}{rgb}{.97, .97, .97}
\definecolor{messagecolor}{rgb}{0, 0, 0}
\definecolor{warningcolor}{rgb}{1, 0, 1}
\definecolor{errorcolor}{rgb}{1, 0, 0}

\usepackage{alltt}

\usepackage[authoryear]{natbib}
\usepackage{times}
\usepackage{amsmath,amsfonts}
\usepackage{graphicx}
\usepackage{setspace}
\usepackage{indentfirst}
\usepackage{url}
\usepackage[font=footnotesize]{caption}

\usepackage{hyperref}
\hypersetup{colorlinks, allcolors={black}}  
\hypersetup{pdfpagemode=UseNone} 

\usepackage[authoryear]{natbib}
\bibpunct{(}{)}{;}{a}{}{,}

\IfFileExists{upquote.sty}{\usepackage{upquote}}{}
\begin{document}



\setstretch{2.0}

\vspace*{8mm}
\begin{center}

\textbf{\Large Identification and correction of sample mix-ups \\[18pt]
in expression genetic data: A case study}

\bigskip \bigskip \bigskip \bigskip

{\large Karl W. Broman$^{*,1}$,
Mark P. Keller$^{\dagger}$,
Aimee Teo Broman$^{*}$, \\[-4pt]
Christina Kendziorski$^{*}$,
Brian S. Yandell$^{\ddagger, \S}$,
\'Saunak Sen$^{**,2}$,
Alan D. Attie$^{\dagger}$}

\bigskip \bigskip

$^{*}$Department of Biostatistics and Medical Informatics,
$^{\dagger}$Department of Biochemistry,
$^{\ddagger}$Department of Statistics,
and $^{\S}$Department of Horticulture,
University of Wisconsin--Madison, Madison, Wisconsin
53706, and
$^{**}$Department of Epidemiology and Biostatistics, University
of California, San Francisco, California 94107
\end{center}

\newpage

\noindent \textbf{Running head:} Correcting sample mix-ups in eQTL data

\bigskip \bigskip \bigskip

\noindent \textbf{Key words:} quality control, microarrays,
genetical genomics, mislabeling errors, eQTL

\bigskip \bigskip \bigskip

\noindent \textbf{$^1$Corresponding author:}

\begin{tabular}{lll}
 \\
 \hspace{1cm} & \multicolumn{2}{l}{Karl W Broman} \\
 & \multicolumn{2}{l}{Department of Biostatistics and Medical Informatics} \\
 & \multicolumn{2}{l}{University of Wisconsin--Madison} \\
 & \multicolumn{2}{l}{2126 Genetics-Biotechnology Center} \\
 & \multicolumn{2}{l}{425 Henry Mall} \\
 & \multicolumn{2}{l}{Madison, WI 53706} \\
 \\
 & Phone: & 608--262--4633 \\
 & Email: & \verb|kbroman@biostat.wisc.edu|
\end{tabular}

\bigskip \bigskip \bigskip

\noindent \textbf{$^2$Present address:} Division of Biostatistics,
Department of Preventive Medicine, University of Tennessee Health
Science Center, Memphis, TN 38163

\newpage

\centerline{\sffamily \textbf{Abstract}}

In a mouse intercross with more than 500 animals and genome-wide gene
expression data on six tissues, we identified a high proportion (18\%)
of sample mix-ups in the genotype data.
Local expression quantitative trait loci (eQTL; genetic
loci influencing gene expression) with extremely large effect were
used to form a classifier to predict an individual's eQTL genotype
based on expression data alone.  By considering multiple eQTL and
their related transcripts, we identified numerous individuals whose
predicted eQTL genotypes (based on their expression data) did not
match their observed genotypes, and then went on to identify other
individuals whose genotypes did match the predicted eQTL genotypes.
The concordance of predictions across six tissues indicated that the
problem was due to mix-ups in the genotypes
(though we further identified a small
number of sample mix-ups in each of the six panels of gene expression
microarrays).
Consideration of the
plate positions of the DNA samples indicated a number of off-by-one and
off-by-two errors, likely the result of pipetting errors.  Such sample
mix-ups can be a problem in any genetic study, but eQTL
data allow us to identify, and even correct, such problems.
Our methods have been implemented in an R package, R/lineup.

\newpage
\centerline{\sffamily \textbf{Introduction}}

To map the genetic loci influencing a complex phenotype, one seeks to
establish an association between genotype and phenotype.  In such an
effort, the maintenance of the concordance between genotyped and
phenotyped samples and data is critical.  Sample mislabelings and
other sample mix-ups will weaken associations, resulting in reduced
power and biased estimates of locus effects.  In traditional genetic
studies, one has limited ability to detect sample mix-ups and almost
no ability to correct such problems.  Inconsistencies between
subjects' sex and X chromosome genotypes may reveal some problems, and
in family studies, some errors may be revealed through Mendelian
inconsistencies at markers, but we will generally be blind to most
errors.

In expression genetics studies, in which genome-wide
gene expression is assayed along with genotypes at genetic markers,
the presence of expression quantitative trait loci (eQTL) with
profound effect on gene expression (particularly local-eQTL, in which
a polymorphism near a gene affects the expression of
that gene) provides an opportunity to not just identify but also
correct sample mix-ups.

In a mouse intercross with more than 500 animals and genome-wide gene
expression data on six tissues, we identified a high proportion (18\%)
of sample mix-ups in the genotype data.  We further identified a small
number of mix-ups among the expression arrays in each tissue.

A number of investigators have developed methods for identifying such
sample mix-ups \citep{Westra2011, Schadt2012, Lynch2012, Ekstrom2012},
and a similar approach was applied by \citet{Baggerly2008,
Baggerly2009} in their forensic bioinformatics analyses of the Duke
debacle.  We have developed a further approach that is simple but
effective.  We illustrate its use through a particularly dramatic
example.

\clearpage

\centerline{\sffamily \textbf{Methods}}

\noindent \textbf{\sffamily Mice and genotyping}
\nopagebreak

C57BL/6J (abbreviated B6 or B) and BTBR \emph{T}$^+$ \emph{tf}/J
(abbreviated BTBR or R) mice were purchased from the Jackson
Laboratory (Bar Harbor, ME) and bred at the University of
Wisconsin--Madison.  The \emph{Lep$^{ob}$\/} mutation was introgressed
into all strains using heterozygous parents to generate homozygous
\emph{Lep$^{ob/ob}$\/} offspring.  F$_2$ mice, all
\emph{Lep$^{ob/ob}$\/}, were the offspring of F$_1$ parents derived
from a cross between BTBR females and B6 males (Figure~S1).
F$_2$ mice and a small number of parental and F$_1$ controls were
genotyped with the 5K GeneChip (Affymetrix).

\bigskip
\noindent \textbf{\sffamily Gene expression microarrays}
\nopagebreak

Gene expression was assayed with custom two-color ink-jet microarrays
manufactured by Agilent Technologies (Palo Alto, CA).
RNA preparations were performed at Rosetta Inpharmatics (Merck \&
Co.). Six tissues were considered: adipose, gastrocnemius muscle
(abbreviated gastroc), hypothalamus (abbreviated hypo), pancreatic
islets (abbreviated islet), kidney, and liver.  Tissue-specific mRNA
pools were used for the second channel, and gene expression was
quantified as the ratio of the mean log$_{10}$ intensity (mlratio).
For further details, see \citet{Keller2008}.

\bigskip
\noindent \textbf{\sffamily Sample mix-ups in the gene expression arrays}
\nopagebreak

\begin{figure}[p]
\centerline{\includegraphics[width=\textwidth]{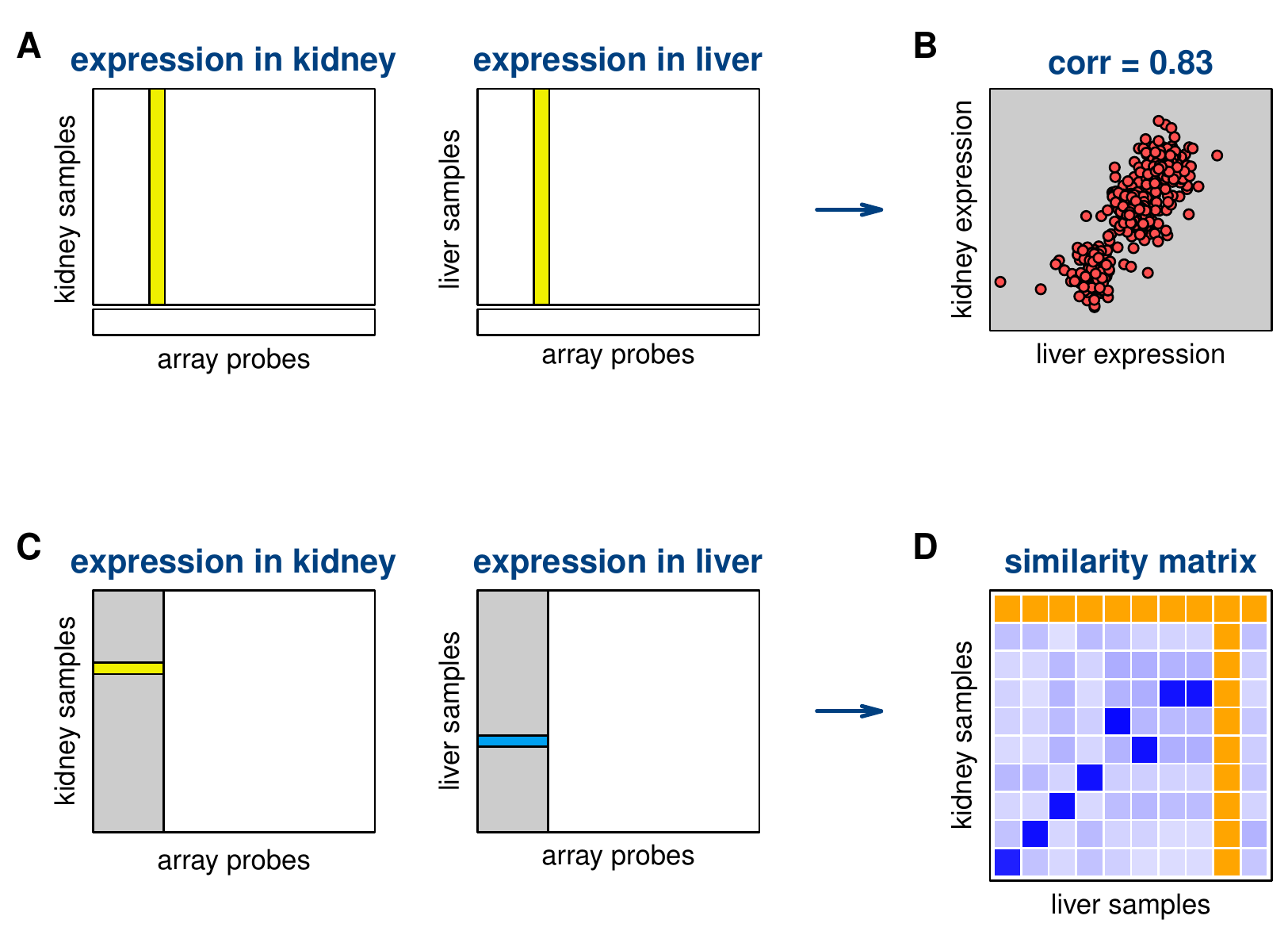}}

\vspace{1cm}

\caption{  Scheme for evaluating the similarity between expression
  arrays for different tissues.  We first consider the expression of
  each array probe for samples assayed for both tissues (A) and calculate
  the between-tissue correlation in expression (B).  We identify the
  subset of array probes with correlation $>$ 0.75 (shaded region in C)
  and calculate the correlation in gene expression for one sample in
  the first tissue and another sample in the second tissue, across
  these selected probes.  This forms a similarity matrix (D), for which
  darker squares indicate greater similarity.  Orange
  squares indicate missing values (samples assayed in one tissue
  but not the other).
\label{fig:eve_scheme}}
\end{figure}

Let $x^s_{ip}$ denote the gene expression measure for sample $i$ at
array probe $p$ in tissue $s$.  We first considered each probe and each
pair of tissues and calculated the between-tissue correlation across
samples, omitting any samples with missing data for that probe in
either tissue.  We identified the subset of probes, for each tissue
pair, with correlation $>$ 0.75.  With this subset of probes, we then
calculated the correlation between sample $i$ in tissue $s$ and sample
$j$ in tissue $t$; call it $r^{st}_{ij}$.  As an illustration,
consider the schematic in Figure~\ref{fig:eve_scheme}: for each pair of tissues, we
identified the subset of probes with high between-tissue correlation
(the shaded region) and then evaluated the correlation between a
sample in one tissue and another sample in the other tissue, across
that subset of probes.

We then summarized the similarity between sample $i$ in tissue $s$
and sample $j$ in the other tissues by the median correlation across
tissue pairs that include tissue $s$,
$r^s_{ij} = \text{median}\{r^{st}_{ij} : t \ne s\}$.  Of course, we
considered only pairs of tissues $(s,t)$ for which sample $i$ was
measured in tissue $s$ and sample $j$ was measured in tissue $t$.

Sample mix-ups in tissue $s$ were identified as samples $i$ for
which the self similarity, $r^s_{ii}$, was small, but for which there
existed some array with high similarity: $\max_{j \ne i} r^s_{ij}$ is
large.  We then inferred the correct label for sample $i$ in tissue
$s$ to be $\arg \max_{j \ne i} r^s_{ij}$.  In other words, viewing
$r^s_{ij}$ as a similarity matrix, we were looking for rows with a
small value on the diagonal, but with some large off-diagonal element
in that row.  In order to ensure confidence in the relabeling of such
samples, we compared the maximum value in the row to the second-highest
value.

To further investigate possible sample duplicates within a tissue,
we considered the subset of probes with correlation $>$ 0.75 with at
least one other tissue, and then calculated between-sample
correlations, across the chosen subset of probes, within that tissue.

\bigskip
\noindent \textbf{\sffamily Sample mix-ups in the DNA samples}
\nopagebreak

In our investigation of potential sample mix-ups in the DNA samples, we
first calculated multipoint genotype probabilities at all markers and
at pseudomarker positions between markers.  The pseudomarker positions
were placed at evenly spaced locations between markers, with a maximum
spacing of 0.5~cM between adjacent markers or pseudomarkers.  The
multipoint genotype probabilities calculations were performed via a
hidden Markov model (HMM), with an assumed genotyping error rate of
0.2\% and with the Carter-Falconer map function \citep{Carter1951}.

We first considered each tissue, individually, and identified the
subset of probes with a strong local-eQTL.  We considered all array
probes with known genomic location and on an autosome, identified the
nearest marker or pseudomarker to the location of the probe, and
calculated a LOD score (log$_{10}$ likelihood ratio) assessing the association between genotype at
that location and the gene expression of that probe.  The LOD score
was calculated by Haley-Knott regression \citep{Haley1992}, a quick
approximation to standard interval mapping \citep{Lander1989}.
Calculations were performed at a single location for each array probe,
rather than with a scan of the genome.  We chose the subset of probes
with LOD $>$ 100.

Continuing to focus on one tissue at a time, we considered the set of
local-eQTL locations and the corresponding probe or probes.
(Generally there was a single probe corresponding to a given eQTL
location, but in a small number of instances for each tissue, there
were a pair of probes at the same eQTL location; for islet, there were
three~eQTL with three corresponding probes, and for
adipose there was one such trio.)  For each eQTL
position and for each mouse, we took the genotypes with maximal
multipoint probability to be the observed eQTL genotype, provided that
this exceeded 0.99; if no genotype had probability $>$ 0.99, the
observed eQTL genotype was treated as missing.

Considering each eQTL in a tissue individually, we then formed a
$k$-nearest neighbor classifier, with $k=40$, for predicting eQTL
genotype from the expression values for the corresponding probe or
probes.  For a given mouse, if more than 80\% of the 40 nearest
neighbors, by Euclidean distance, shared the same observed eQTL
genotype, this was taken to be the inferred eQTL genotype for that mouse.
If no more than 80\% of the 40 nearest neighbors shared a common
genotype, the inferred eQTL genotype was treated as missing.

In order to filter out samples that were clearly incorrect and improve
our classifiers, we then calculated the proportion of matches, for
each sample, between the observed eQTL genotypes and the corresponding
inferred eQTL genotypes, omitted samples for which the proportion of
matches was $<$ 0.7, and rederived the $k$-nearest neighbor classifiers
with the subset of samples deemed likely correct.

\begin{figure}[p]
\centerline{\includegraphics[width=\textwidth]{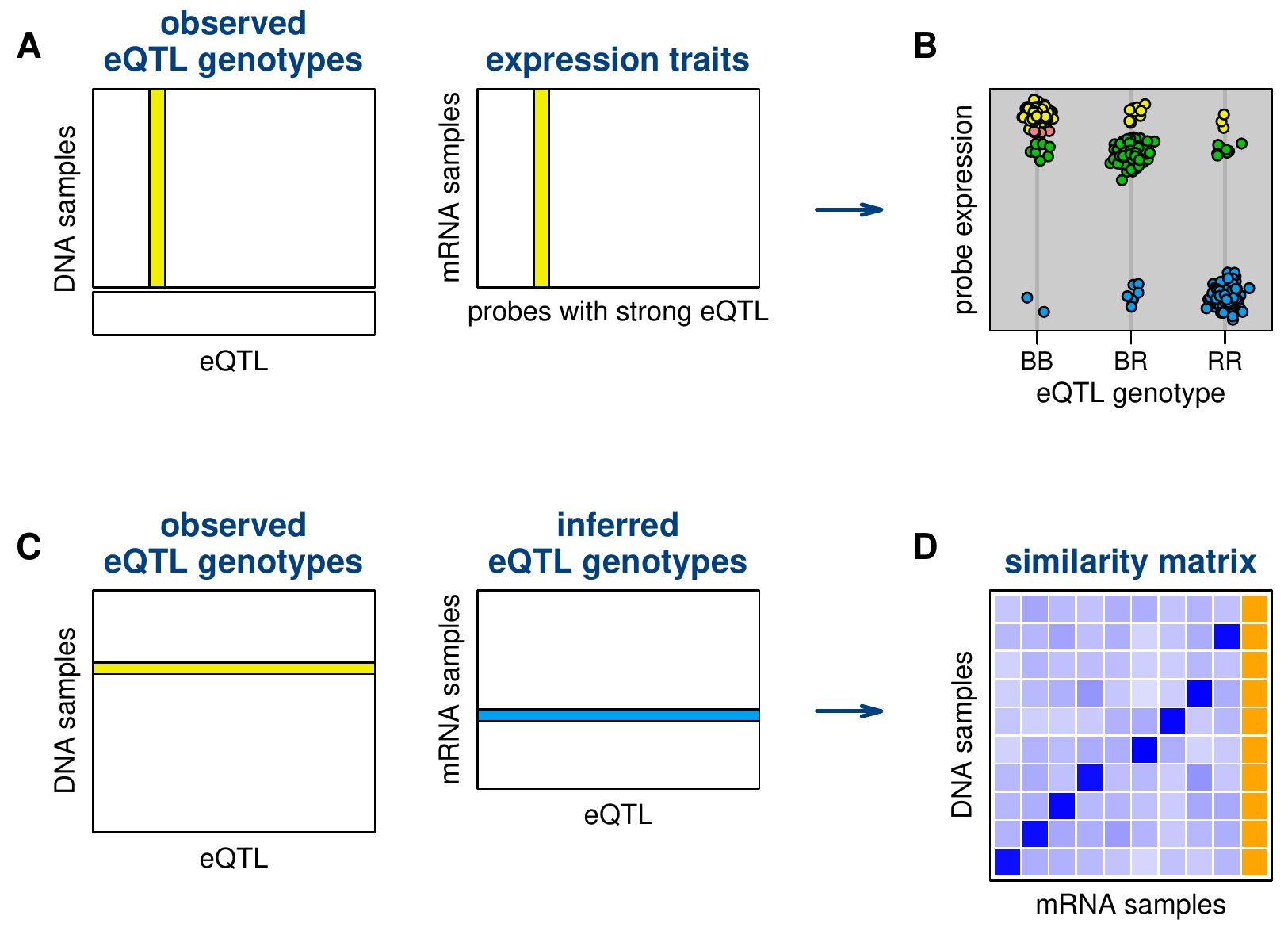}}

\vspace{1cm}

\caption{  Scheme for evaluating the similarity between genotypes and expression
  arrays.  We first identify a set of probes with
  strong local eQTL.  For each such eQTL, we use the samples with both
  genotype and expression data (A) to form a classifier for predicting
  eQTL genotype from the expression value (B).  We then compare the
  observed eQTL genotypes for one sample to the inferred eQTL
  genotypes, from the classifiers, for another sample (C).  The
  proportion of matches, between the observed and inferred genotypes,
  forms a similarity matrix (D), for which darker squares indicate
  greater similarity.  Orange squares indicate missing values (for
  example, samples with genotype data but no expression data).
\label{fig:gve_scheme}}
\end{figure}

As an illustration, consider the schematic in Figure~\ref{fig:gve_scheme}: for each
tissue, we identified a subset of array probes with strong local-eQTL,
we derived classifiers for predicting eQTL genotype from the
corresponding expression phenotypes, and then constructed a matrix of
inferred eQTL genotypes.  As a measure of similarity between a DNA
sample and an mRNA sample, we calculated the proportion of matches
between the observed eQTL genotypes for the DNA sample and the
inferred eQTL genotypes for the mRNA sample.

To combine the tissue-specific similarity measures across the six
tissues, we simply took the overall proportion of matching genotypes,
across all eQTL and across all tissues.

As in the investigation of sample mix-ups within the expression
arrays, we treated the proportions of matches between observed and inferred eQTL
genotypes as a similarity matrix.  Problem DNA samples were identified
as rows for which the value on the diagonal (the self similarity) was
small.  In such rows, we inferred the correct label to be that of the
maximal off-diagonal value, provided that this maximum was large and
was well above the second-largest value.

\bigskip
\noindent \textbf{\sffamily QTL analysis}
\nopagebreak

To characterize the improvement in results following correction of
sample mix-ups, we performed QTL analysis with several traits of
interest, including the expression traits in each tissue, with the
original data and with the corrected data.  In the corrected data, we
omitted the DNA samples that could not be verified to be correct
(that is, those with no corresponding gene expression data.)

\bigskip
\noindent \textbf{\emph{Insulin:}} We first considered a clinical phenotype
of considerable interest: 10 week plasma insulin.  QTL analysis was
performed by Haley-Knott regression \citep{Haley1992}, with log
insulin, and with sex included as an interactive covariate (that is,
allowing the effects of QTL to be different in the two sexes).

\bigskip
\noindent \textbf{\emph{Agouti and tufted coat:}} We considered two simple
Mendelian traits: agouti coat color (due to a single gene on
chromosome 2) and tufted coat (due to a single gene on chromosome 17).
QTL analysis was performed treating each phenotype as a binary trait
\citep{Xu1996, Broman2003b}.  To handle possible marker genotyping
errors at the causal loci, we took the observed genotypes to be those
with maximal multipoint probability, provided that this exceeded 0.99;
if no genotype had probability $>$ 0.99, the observed genotype was
treated as missing.


\bigskip
\noindent \textbf{\emph{eQTL analyses:}} We considered each of the six
tissues individually, and focused on the subset of probes with known
genomic location on an autosome or the X chromosome.  For
hypothalamus tissue, we omitted a batch of 119 poorly behaved arrays,
though these had been included in our efforts to identify sample mix-ups.

Expression measures were transformed to normal quantiles.  That is,
the expression measures were converted to ranks $R_i \in \{1, \dots,
n\}$ and then transformed to $y_i = \Phi^{-1}[(R_i - 0.5)/n]$, where
$\Phi^{-1}$ is the inverse of the normal cumulative distribution
function.

QTL analysis was performed by Haley-Knott regressions with
sex included as an interactive covariate.  We considered the maximal
peak for each array probe on each chromosome, and inferred the
presence of a QTL if the LOD score exceeded 5, a 5\% genome-wide
significance level established by computer simulation.

An inferred eQTL was considered a local-eQTL if the
2-LOD support interval contained the genomic location of the
corresponding array probe; otherwise, it was considered a
\emph{trans}-eQTL.

\bigskip
\noindent \textbf{\sffamily Software}
\nopagebreak

All analyses were conducted with R \citep{R}.  QTL analyses were
performed with the R package, R/qtl \citep{Broman2003a}.  Our methods
for identifying sample mix-ups have been assembled as an R package,
R/lineup, available at \url{http://github.com/kbroman/lineup} as well
as The Comprehensive R Archive Network (CRAN;
\url{http://cran.r-project.org}).

\bigskip
\noindent \textbf{\sffamily Data availability}
\nopagebreak

The genotype and gene expression microarray data are available at the
QTL Archive, now part of the Mouse Phenome Database:

\noindent
\url{http://phenome.jax.org/db/q?rtn=projects/projdet&reqprojid=532}

\clearpage

\centerline{\sffamily \textbf{Results}}

We first became aware of potential problems in the samples through the
identification of six duplicate DNA samples and
32~mice whose X
chromosome genotypes were incompatible with their sex.
We genotyped 554~F$_2$ mice at
2060~informative SNPs, including 20
on the X chromosome.  Three samples were assigned ``no call'' at all
markers and not considered further.  Six pairs were seen to be
duplicates, with over 98\% genotype identity across typed markers
(Table~S1).

The F$_2$ mice were the offspring of F$_1$ siblings derived by
crossing BTBR females to B6 males (Figure~S1).  F$_2$ females should
be homozygous BTBR (RR) or heterozygous (BR) on the X chromosome;
F$_2$ males should be hemizygous B or R.  (Note that homozygous and hemizygous
genotypes could not be distinguished.)  However,
19~females exhibited some homozygous B6
genotypes on the X, and 17~males exhibited some
heterozygous genotypes (Figure~S2).  While four of these males had a
single heterozygous genotype that was likely a genotyping error, the
19~females and the other
13~males were clearly indicated to have swapped
sex.  There were an additional 53~females and
50~males with homozygous RR or hemizygous R genotypes for
all markers on the X chromosome, compatible with either sex.

In cleaning the genotype data, we omitted a set of
seven samples, including one pair of the sample
duplicates, with poorly behaved data.  (They showed a high rate of
apparent genotyping errors, an unusually large proportion of
homozygous genotypes, and an unusually large number of apparent
crossovers.)  For the other five pairs of duplicates, we omitted one
sample from each pair.

\bigskip
\noindent \textbf{\sffamily Sample mix-ups in the gene expression arrays}
\nopagebreak


For each of six tissues
(adipose, gastroc, hypo, islet, kidney, liver),
approximately 500~F$_2$ mice were assayed for gene expression with
two-color Agilent arrays with tissue-specific pools (Table~S2).  A
small number of poorly behaved arrays were omitted.  We later
discovered a batch of 119 poorly behaved arrays for
hypo, but these were included in the analyses described here.
There were 527~mice assayed for at least one of
the six tissues,
but not all mice were assayed for all tissues.
In particular, there were 27~mice assayed only for
gene expression in kidney, and 43~mice assayed for
all tissues except kidney.  Further, 27~mice were genotyped
but were not subject to gene expression analysis.

To identify potential sample mix-ups among gene expression arrays, we
first identified, for each
pair of tissues, a subset of array probes with high between-tissue
correlations.  Consideration of all probes would greatly reduce the
apparent correlation between arrays, due to the abundance of
unexpressed genes.  For example, for Mouse3567, the correlation
between gene expression in kidney and in liver, across all
40,572
probes, is 0.32, while for the subset of
155~probes with correlation $>$ 0.75 between kidney and
liver, the correlation is 0.78.  (See Figure~S3.)

Figure~S4 contains density estimates of the between-tissue
correlations for all array probes.  The densities are organized by
tissue, with the panel for each tissue containing the five tissue pairs
involving that tissue.  There are some small differences among tissue
pairs, but the vast majority of between-tissue correlations are between
-0.25 and 0.50.  Table~S3 contains the numbers of probes for each pair
of tissues with correlations exceeding 0.70, 0.75, 0.80, and 0.90,
respectively.  We focused on probes with correlations $>$ 0.75, of
which there were between 46 and
200 probes per tissue pair.

For each pair of tissues, we calculated the correlations among
samples across the subset of correlated probes.  For each tissue, we
then summarized the similarity between each sample in that tissue and
each sample in other tissues by the median correlations, across the
tissue pairs that included the target tissue.

Figure~S5 contains histograms of the similarity measures for each
tissue, separating the self-self similarities (the diagonal elements)
and the self-nonself similarities (the off-diagonal elements).  There
are a number of clear outliers: small self-self similarities and large
self-nonself similarities.  The self-nonself similarities follow a
bimodal distribution, with the lower mode corresponding to
opposite-sex pairs and the upper mode corresponding to same-sex
pairs.  The chosen probes included a probe in \emph{Xist} (involved in
X chromosome inactivation) and probes on the Y chromosome.

\begin{figure}

\centerline{\includegraphics[width=6.05in]{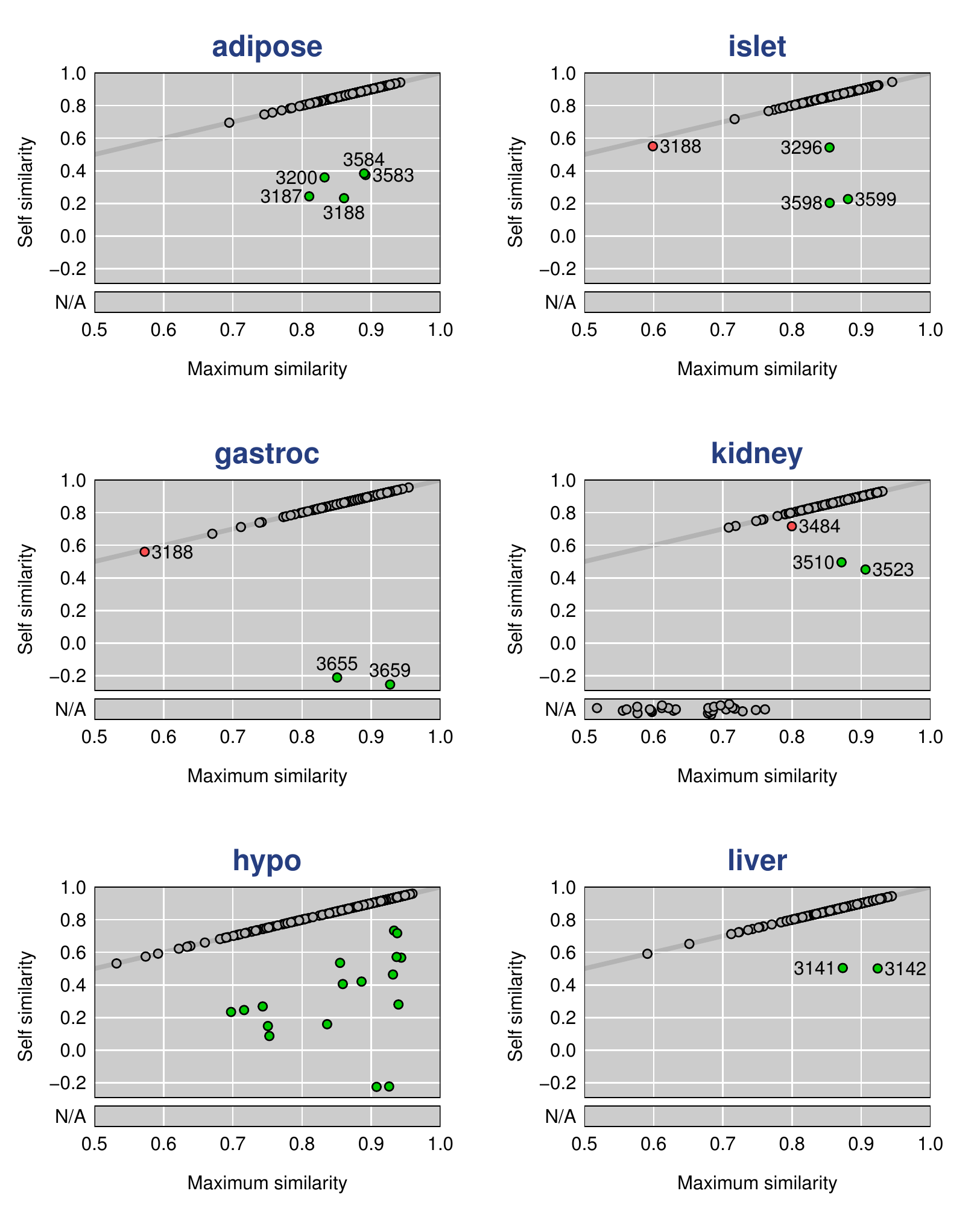}}

\vspace*{4mm}

\caption{  Self similarity (median correlation across tissue pairs) versus
  maximum similarity for the expression arrays for each tissue.  The
  diagonal gray line corresponds to equality.  Green points are
  inferred to be sample mix-ups. Gray points correspond to arrays for
  which the self similarity is maximal.  Red points correspond to
  special cases (see the text).  There were
  27~samples assayed only for kidney; these have
  missing self similarity values.
\label{fig:eve_similarity}}
\end{figure}

To identify problem samples in each tissue, we considered for each
sample, the self similarity vs.\ the maximum similarity (that is, the
values on the diagonal of the similarity matrix and the maximum values
in each row).  These are displayed in Figure~\ref{fig:eve_similarity}.

The vast majority of samples in each tissue were indicated to be
correctly labeled: the self similarity was the maximum similarity.
But for each tissue, there were at least a few samples which were more
like some other sample in the other tissues.  In each case, we infer
the correct label to be that with the maximal similarity.  In
Figure~S6, we display the second-highest similarity vs.\ the maximum
similarity for each sample in each tissue.  The problem
samples (colored green) are generally well away from the diagonal,
indicating good support for our ability to infer the correct label.

The red points in Figure~\ref{fig:eve_similarity} and Figure~S6 are
special cases: The Mouse3188 sample is highlighted as a potential problem in both
islet and gastroc (being slightly off the diagonal line), but this is
because that sample was involved in array swaps in two different
tissues (adipose and hypo).  This is the only sample indicated to be
mislabeled in multiple tissues.  We also highlight Mouse3484 in
gastroc, which appeared to be a mixture (more on this below).

\begin{figure}

\centerline{\includegraphics[width=\textwidth]{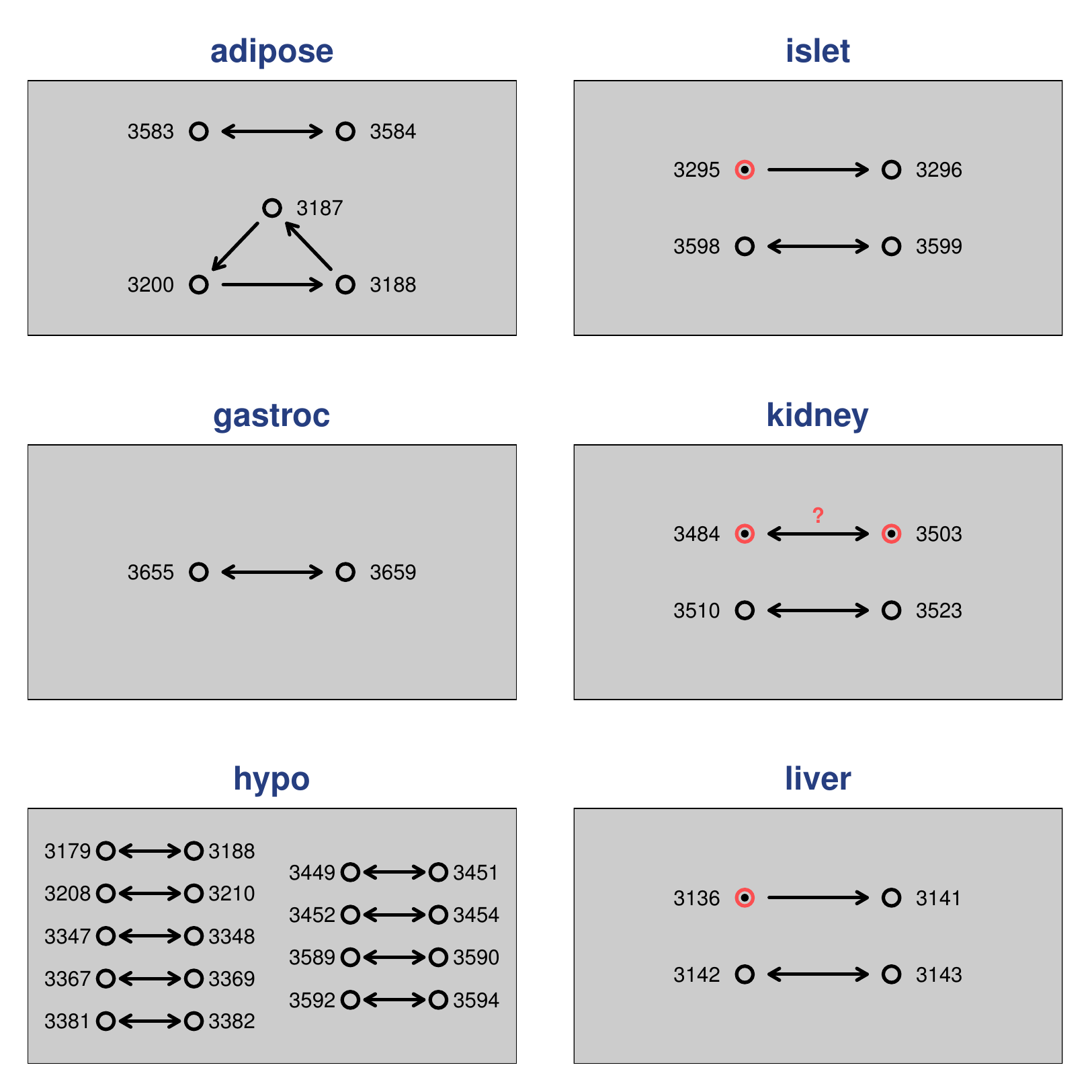}}

\vspace{1cm}

\caption{  The mRNA sample mix-ups for the six tissues.  Double-headed arrows
  indicate a sample swap.  The trio of points in adipose corresponds
  to a three-way swap.  The pink circles with a single-headed arrow,
  in islet and liver, are sample duplicates.  The questionable case in
  kidney indicates a potential sample mixture arrayed twice.
\label{fig:expr_swaps}}
\end{figure}

The inferred errors are displayed in Figure~\ref{fig:expr_swaps}.
For adipose, we identified two problems.  The samples for Mouse3583
and Mouse3584 were swapped, and there was a three-way swap among
Mouse3187, Mouse3188, and Mouse3200, with the sample labeled Mouse3187
really being Mouse3188, that labeled Mouse3188 really being Mouse3200,
and that labeled Mouse3200 really being Mouse3187.

For gastroc, there was a single sample swap, between Mouse3655 and
Mouse3659.  For hypo, there were 9 pairs of sample swaps.  For islet,
the samples Mouse3598 and Mouse3599 were swapped, and the sample
labeled Mouse3296 was really a duplicate (or \emph{unintended
  technical replicate}) of the Mouse3295 sample.  For liver, the
sample labeled Mouse3142 really corresponded to Mouse3143 (Mouse3142
was not assayed for gene expression in liver), and the sample labeled
Mouse3141 was really a duplicate of the Mouse3136 sample.

For kidney, the samples for Mouse3510 and Mouse3523 were swapped, and
Mouse3484 was also seen to be a problem.  We believe that the samples for
Mouse3484 and Mouse3503 may have been mixed and assayed twice in
duplicate (more below).  There were 27~samples that
were assayed for gene expression only in kidney; for these, the self
similarity cannot be calculated.  We have limited ability to detect
mix-ups for these samples, but none were very close to any sample
in other tissues, and so they can, at least provisionally, be assumed
to be correctly labeled.

To further illustrate the sample swaps, Figure~S7 contains scatter
plots of the gastroc arrays labeled Mouse3655 and Mouse3659 against
the arrays in the other tissues with those labels.  For each pair of
tissues, we plot the array probes with between-tissue correlation $>$
0.75.  Mouse3655 in gastroc is correlated with Mouse3659 in other
tissues, while Mouse3659 in gastroc is correlated with Mouse3655 in
other tissues, indicating a clear swap between these samples within
gastroc.

Figure~S8 contains similar scatter plots for a pair of inferred
duplicates, with the sample labeled Mouse3141 in liver really being a
duplicate of the Mouse3136 liver sample.  Mouse3136 liver and
Mouse3141 liver are each correlated with Mouse3136 in other tissues
and not with Mouse3141, and the two samples are extremely highly
correlated with each other (see the two central panels in the bottom
row).  In Figure~S9, we display the between-sample correlations for
samples with these two labels, for all pairs of tissues, with the
pairs including liver highlighted in red.  The Mouse3136 samples are
correlated for all tissue pairs; the Mouse3141 samples are correlated
for all tissue pairs not involving liver, and the Mouse3141 liver
sample is correlated with all Mouse3136 samples in other tissues.

The Mouse3484 and Mouse3503 samples in kidney appear to be sample
duplicates, but these samples are correlated with each of Mouse3484
and Mouse3503 in the other tissues.  We're inclined to believe that
the two kidney samples were mixed and arrayed in duplicate, but we are
not able to prove this point.  Figure~S10 contains scatter plots for
the two samples in kidney vs.\ all tissues; the central panels in the
second row from the bottom indicate that the two samples are highly
correlated and so likely replicates, but all scatter plots here show
strong correlation.  Figure~S11 contains the between-sample
correlations for both sample labels in all tissue pairs; contrast this
with Figure~S9, for the simple duplicate in liver.  Mouse3484 kidney
and Mouse3503 kidney are strongly correlated with both samples in
the other tissues, but not so strongly as Mouse3484 and Mouse3503 are
with themselves in the non-kidney pairs.  And for tissue pairs not
including kidney, Mouse3484 and Mouse3503 are much more weakly
correlated.

As we were unable to resolve the problems with Mouse3484 and Mouse3503
in kidney, these two arrays were omitted from later analyses.  The two
simple sample duplicates, one in islet and one in liver, were combined
and assigned the correct label.  The other sample mix-ups were
relabeled as inferred in Figure~\ref{fig:expr_swaps}.

Expression of the \emph{Xist\/} gene (involved in X chromosome inactivation and
so highly expressed in females but not males) and of genes on the Y
chromosome is a useful diagnostic for the sex of an mRNA sample.  In
Figure~S12, we display, for each tissue, the average expression across
for Y chromosome genes vs.\ the expression of \emph{Xist}, with the original
data and after correction of the sample mix-ups in the expression
arrays.  Just three of the sample-swaps (one in gastroc and two in
hypo) involved opposite-sex pairs.  These show up clearly in the left
column, with the original data, and are resolved after correction of
the sample mix-ups.  The unusual pattern of expression in hypo,
with a bimodal distribution for the Y chromosome genes in males and a
large number of females with relatively low \emph{Xist\/} expression,
was due to a set of 119 poorly behaved arrays.

\bigskip
\noindent \textbf{\sffamily Sample mix-ups in the genotypes}
\nopagebreak

Having corrected the sample mix-ups among the gene expression arrays,
we turned to potential problems in the genotypes.  For each tissue,
we considered the 36,364
autosomal array probes with known genomic location and identified
those with a strong local-eQTL, having LOD score $>$ 100 for
the association between the probe expression measures and
genotype at the corresponding location.

\begin{figure}

\centerline{\includegraphics[width=\textwidth]{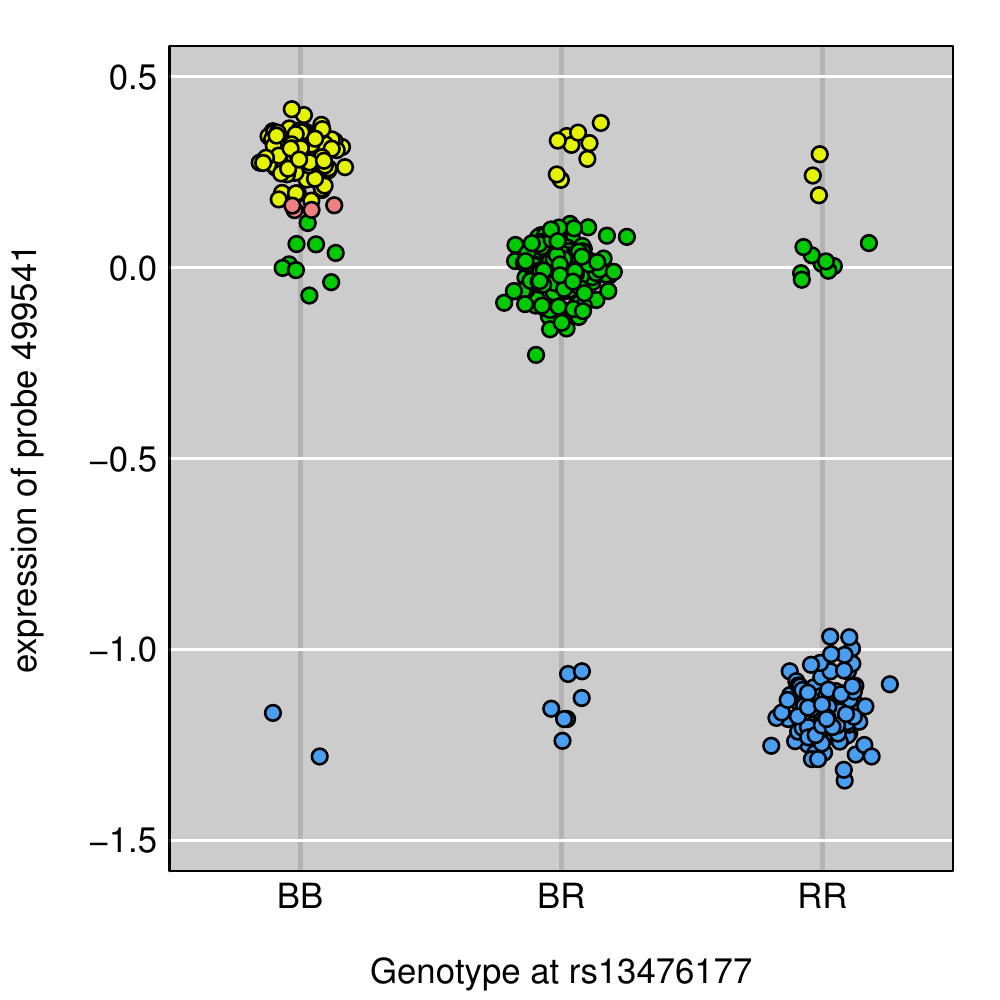}}

\vspace{1cm}

\caption{  Plot of islet expression vs observed genotype for an example
  probe.  Points are colored by the inferred genotype, based on a
  k-nearest neighbor classifier, with yellow, green, and blue
  corresponding to BB, BR, and RR, respectively, where B = B6 and R =
  BTBR.  Salmon-colored points lie at the boundary between two
  clusters and were not assigned.
\label{fig:gve}}
\end{figure}

For each such probe, we created a $k$-nearest neighbor classifier (with
k=40), for predicting eQTL genotype from the expression phenotype.
For example, in Figure~\ref{fig:gve}, we display the expression, in
islet, of probe 499541 (on
chromosome~1) vs.\ genotype
at the nearest marker.  At this probe, there are three clear groups of
mice, with B6 homozygotes B6 (BB) having high expression, BTBR
homozygotes (RR) having low expression, and heterozygotes (BR)
intermediate.  There are a number of mice whose expression does not
match their observed eQTL genotype; the classifier infers a different
eQTL genotype.  The points highlighted in pink have expression at the
boundary between the BB and BR groups and are left unassigned.  (To
assign an inferred eQTL genotype to a point, we required that 80\% of
the nearest neighbors had a common eQTL genotype.)

For sets of probes mapping to approximately the same genomic location,
we considered the probes' expression jointly.  Examples of pairs of
probes mapping to the same location are shown in Figure~S13, with
points colored by observed eQTL genotype.

We considered 45--115 eQTL per tissue;
their locations on the genetic map of markers is shown in Figure~S14.
The majority of eQTL had a single corresponding probe.  There were
3--14 eQTL per tissue with a
pair of corresponding probes.  For islet, there were
three eQTL with three corresponding probes, and for
adipose there was one such trio.

For each tissue, we calculated the proportion of matches between
the observed eQTL genotypes for each DNA sample and the inferred eQTL
genotypes from each mRNA sample, as a measure of similarity between
the DNA and mRNA samples.  We further calculated a combined measure of
similarity as the overall proportion of mismatches, pooling all six
tissues.

Figure~S15 contains histograms of the similarity measures for each
tissue, separating the self-self similarities (the diagonal elements)
and the self-nonself similarities (the off-diagonal elements).  There
are a number of clear outliers: small self-self similarities and large
self-nonself similarities.

\begin{figure}

\centerline{\includegraphics[width=\textwidth]{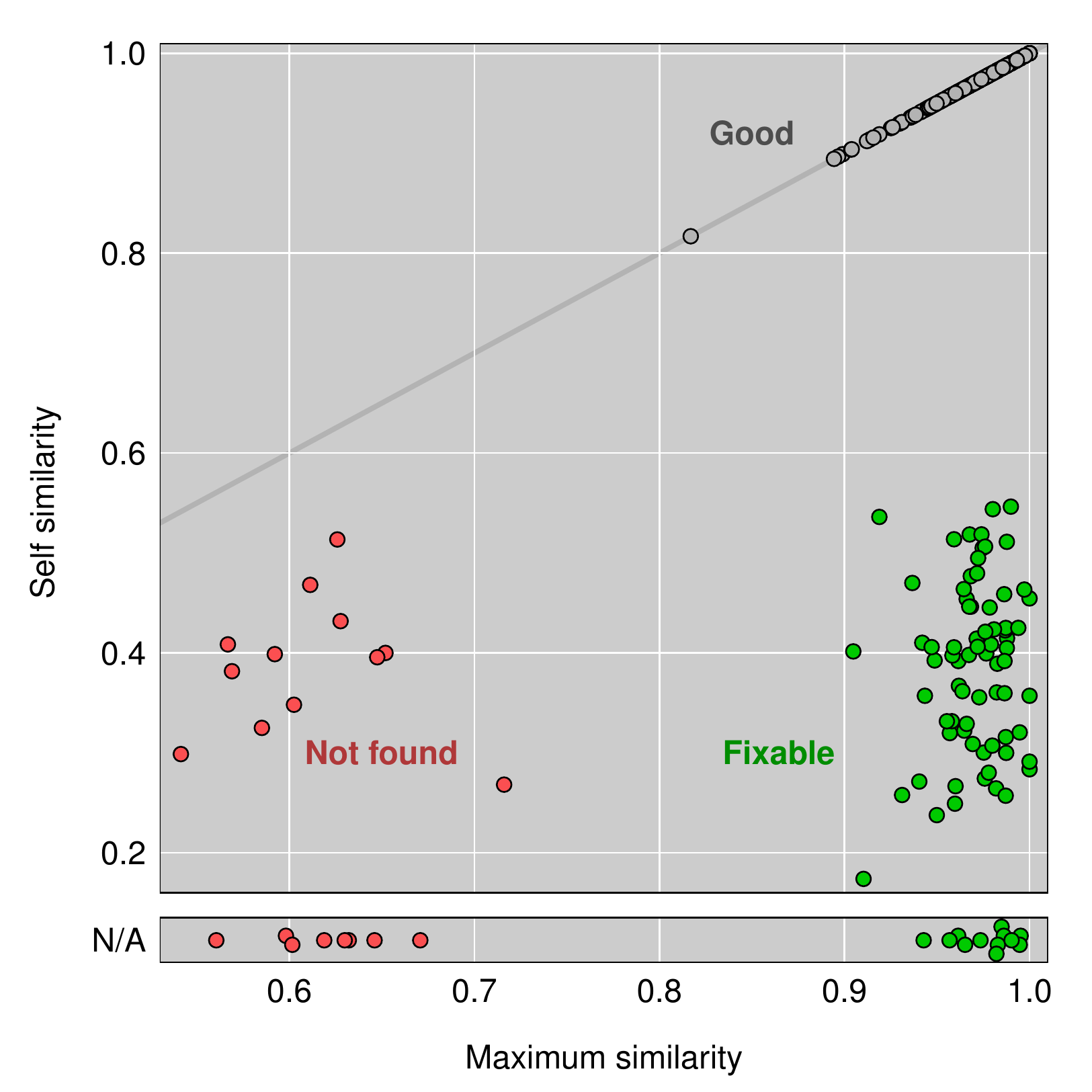}}

\vspace{1cm}

\caption{  Self similarity (proportion matches between observed and
  inferred eQTL genotypes, combined across tissues) versus maximum
  similarity for the DNA samples.  The diagonal gray line corresponds to
  equality.  Samples with missing self similarity (at bottom) were not
  intended to have expression assays performed.  Gray points
  correspond to DNA samples that were correctly labeled.  Green points
  correspond to sample mix-ups that are fixable (the correct label can
  be determined).  Red points comprise both samples mix-ups that
  cannot be corrected as well as samples that may be correct but
  cannot be verified as no expression data is
  available.
\label{fig:gve_similarity}}

\end{figure}

To identify problem
DNA samples,  we again considered the self similarity
vs.\ the maximum similarity (that is, the values on the
diagonal of the similarity matrix vs. the maximum values in each
row).  Figure~\ref{fig:gve_similarity} contains a scatterplot of these
values.  Gray points, with maximum similarity equal to the self
similarity, are inferred to be corrected labeled.  Green points, with
small self similarity but large maximum similarity, are inferred to be
incorrect, but are fixable.  Red points concern DNA samples for which no
corresponding mRNA sample can be found.

Detailed results for the six tissues, with tissue-specific similarity
values, are shown in Figure~S16.  The points are colored as in
Figure~\ref{fig:gve_similarity}, based on the combined similarity
measure.  The points with missing self similarity (at the bottom of
each panel) were not intended to be assayed for gene expression in
that tissue.  The tissue-specific results are concordant with the overall
conclusions, with two caveats.  First, there are a number of green
points (corresponding to mislabeled, but fixable, DNA samples), with
low maximum similarity in each tissue.  These correspond to samples
for which gene expression assays were not performed for that
tissue, the bulk of which are for the 27 samples that
were assayed only for gene expression in kidney and the
43 samples that were assayed for all tissues except
kidney.  Second, for hypo, the strength of eQTL associations were
weaker, and fewer eQTL were considered, than for the other tissues,
and so there is less separation between the green and pink points.

In Figure~S17, we display the second-highest similarity vs.\ the
maximum similarity, for the combined similarity measures accounting
for all tissues.  The fixable mis-labeled samples (in green) are all
well away from the diagonal, indicating good support for our ability
to infer the correct label.

\begin{figure}

\vspace*{-0.25in}

\centerline{\includegraphics[width=5.83in]{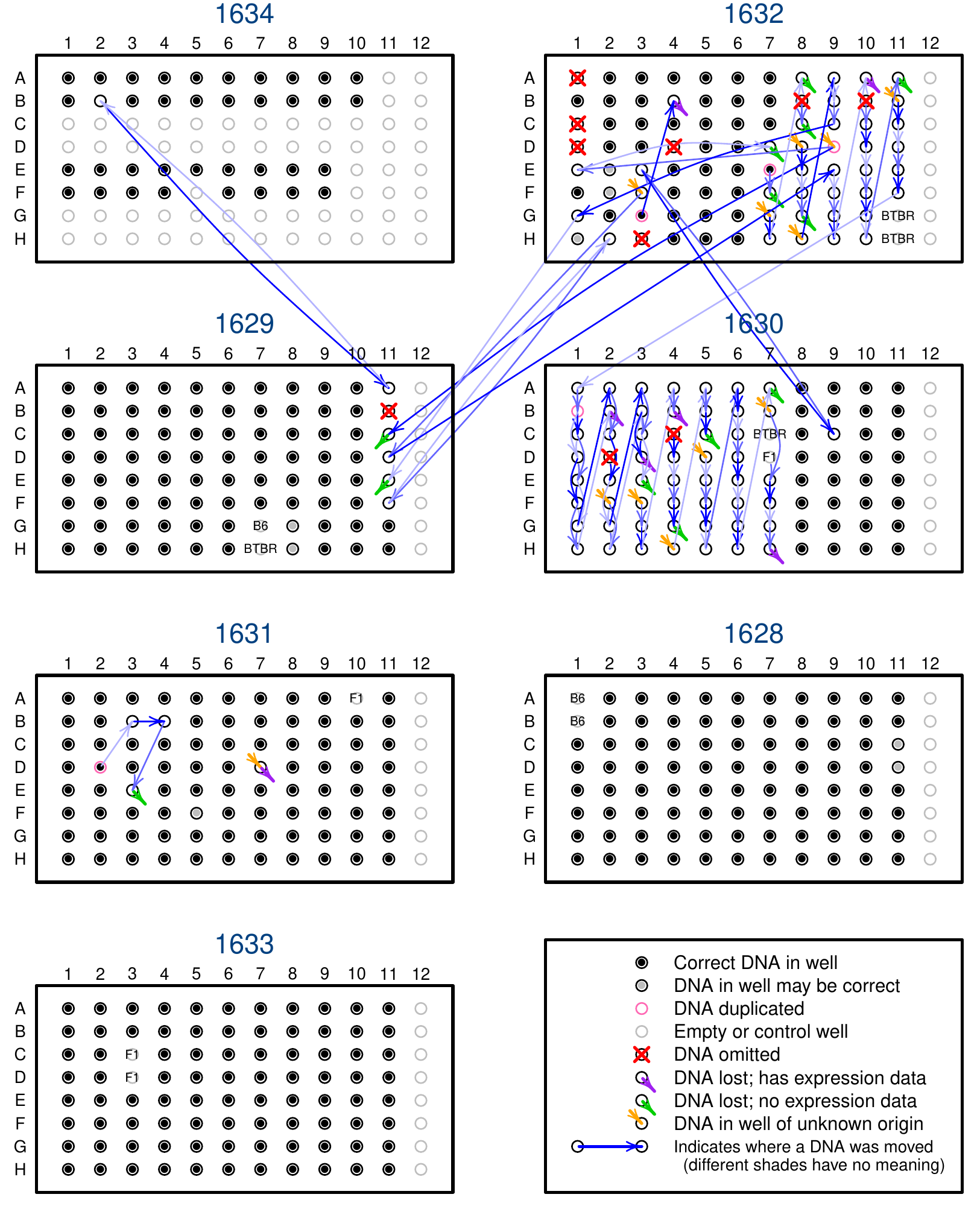}}

\vspace*{4mm}

\caption{\footnotesize   The DNA sample mix-ups on the seven 96-well plates used for
  genotyping.  Black dots indicate that the correct DNA was put in the
  well.  Blue arrows point from where a sample should have been placed to
  where it was actually placed; the different shades of blue convey no
  meaning.  Red X's indicate DNA samples that were omitted.  Orange
  arrowheads indicate wells with incorrect samples, but the sample
  placed there is of unknown origin.  Purple and green arrow-heads
  indicate cases where the sample placed in the well was incorrect,
  but the DNA that was supposed to be there was not found; with the
  purple cases, there was corresponding gene expression data, while
  for the green cases, there was no corresponding gene expression
  data.  Pink circles (e.g., well D02 on plate 1631) indicate sample
  duplicates.  Gray dots indicate that the sample placed in the well
  cannot be verified, as there was no corresponding gene expression
  data.  Gray circles indicate controls or unused wells.
 \label{fig:genotype_plates}}

\end{figure}

The inferred mix-ups among the DNA samples are displayed in
Figure~\ref{fig:genotype_plates} according to the arrangement of the
samples on the 96-well genotyping plates.  Black dots indicate that
the correct DNA sample was placed in the correct well.  The blue arrows point
from the well in which a DNA sample was supposed to be placed, to the
well where it was actually placed.
For example, on plate 1631, the sample in well D02 was placed in the
correct well but was also placed in well B03.  The sample that
belonged in B03 was placed in B04, the sample that belonged in
B04 was placed in E03, and the sample belonging in E03 was not found
(but, as indicated by the green arrowhead, there was no corresponding gene
expression data).

While there were many long-range sample swaps, particularly for
samples belonging in the eleventh column of plate 1629, the bulk of
the errors occurred on plates 1632 and 1630, with a long series of
off-by-one and off-by-two errors indicative of single-channel
pipetting mistakes.

Let us describe a small portion of the further errors.
On plate 1632, the sample belonging in well E07 was placed in
the correct well but was also placed in the well below, F07.  The
sample belonging in well F07 was not found but had no corresponding
gene expression data.  The sample placed in well G07 was incorrect but
had no corresponding gene expression data, and so presumably
corresponds to that which should have been in the well above, F07.
The sample belonging in well G07 was placed one below, H07.  There are
then a series of off-by-one errors, except that the sample belonging
in well C09 was actually placed in well G01, while the sample
belonging in well D09 was placed in both well E01 and on plate 1629
(well C11).

Of the 554 DNA samples that were genotyped, 10
were omitted due to poorly behaved genotypes (including a pair of
replicates), 435 were found to be correctly labeled, and
8 were possibly correct but could not be verified due
to lack of gene expression assays.  However, 5 samples
were duplicates of other samples, 84 were incorrectly
labeled but the correct label could be assigned, and 12
were incorrectly labeled and the correct label could not be
identified.  Thus, at least
18\%
of the samples were involved in sample mix-ups.

We had initially become suspicious of possible sample mix-ups through
the identification of 36~mice
whose X chromosome genotypes were inconsistent with their sex.  After
correction of the sample mix-ups, there were no such discrepancies.
Only a small portion of the problems were identified through such
sex/genotype incompatibilities, because the majority of sample mix-ups
were off-by-one errors in the genotype plates, and the samples were
arranged on the plates so that adjacent samples were often the same
sex.

The large discrepancies between expression and eQTL genotype seen in
Figure~5 and Figure~S13 are largely eliminated following correction of
the inferred sample mix-ups. Figure~S18 shows the same examples, but
with the corrected data. Panels A-D of Figure~S18 correspond to the
panels in Figure~S13; the genotypes are now more clearly separated,
though some overlap remains and there are a few outliers (most
notably, in Figure~S18B). Panel E of Figure~S18 corresponds to
Figure~5; following correction of the sample mix-ups, there is no
overlap between the three genotype groups.

\bigskip
\noindent \textbf{\sffamily QTL mapping results}
\nopagebreak

\begin{figure}

\centerline{\includegraphics[width=\textwidth]{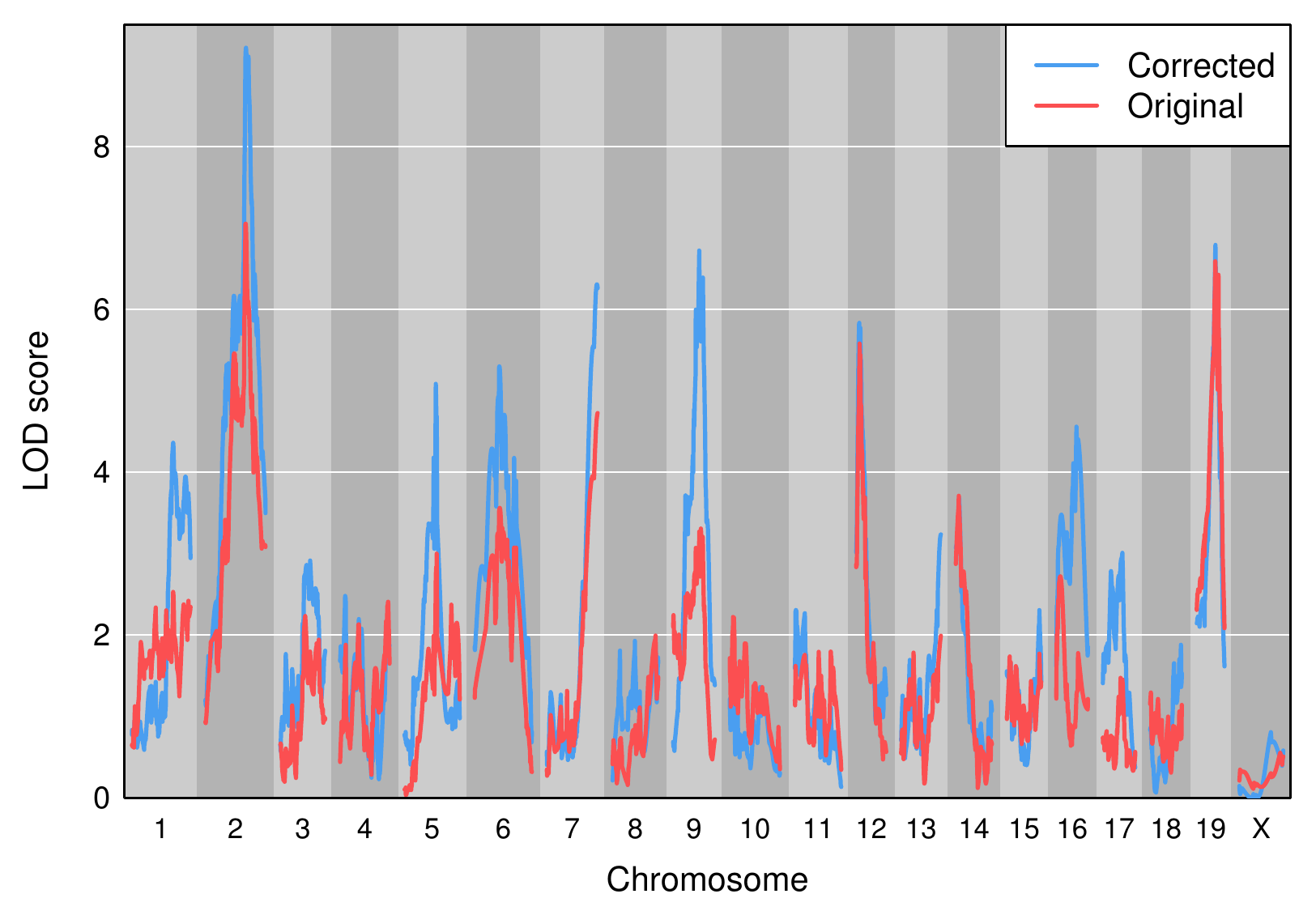}}

\vspace{1cm}

\caption{  LOD curves for 10 week insulin level, before (salmon color) and
  after (blue) correction of the sample mix-ups.
\label{fig:insulin_lod}}
\end{figure}

It should come as no surprise that the correction of the sample
mix-ups, particularly the
18\%
mix-ups in the DNA samples, leads to great improvement in QTL mapping
results.  Figure~\ref{fig:insulin_lod} contains LOD curves for
10~week insulin level with the original and corrected datasets.
With the original data, four chromosomes had LOD score $>$ 4; after
correction of the sample mix-ups, nine chromosomes have LOD score $>$
4.

Two coat-related traits were recorded for the F$_2$ mice: agouti and
tufted coats.  Concerning agouti coat: BTBR mice have tan coats, while
B6 mice are black; this is due to a gene on chromosome 2, and the BTBR
allele is dominant.  Mapping the agouti coat color as a binary
phenotype, the LOD score on chromosome 2 increased from
64 to
110 after correction of the sample
mix-ups (Figure~S19A).  While the corrected data still contained
inconsistencies between genotype and coat color, the number of
inconsistencies decreased from 47 to
7 (Table~S4).

Tufted coat is due to a single gene on chromosome 17, with the BTBR
allele (with the tufted phenotype) being recessive to the B6 allele
(not tufted).  Mapping this phenotype as a binary trait, the LOD score
on chromosome 17 increased from 64 to
107 after correction of the sample
mix-ups (Figure~S19B).  While, as with agouti, the corrected data still contained
inconsistencies between genotype and phenotype, the number of
inconsistencies decreased from 37 to
4 (Table~S5).

\begin{figure}

\centerline{\includegraphics[width=\textwidth]{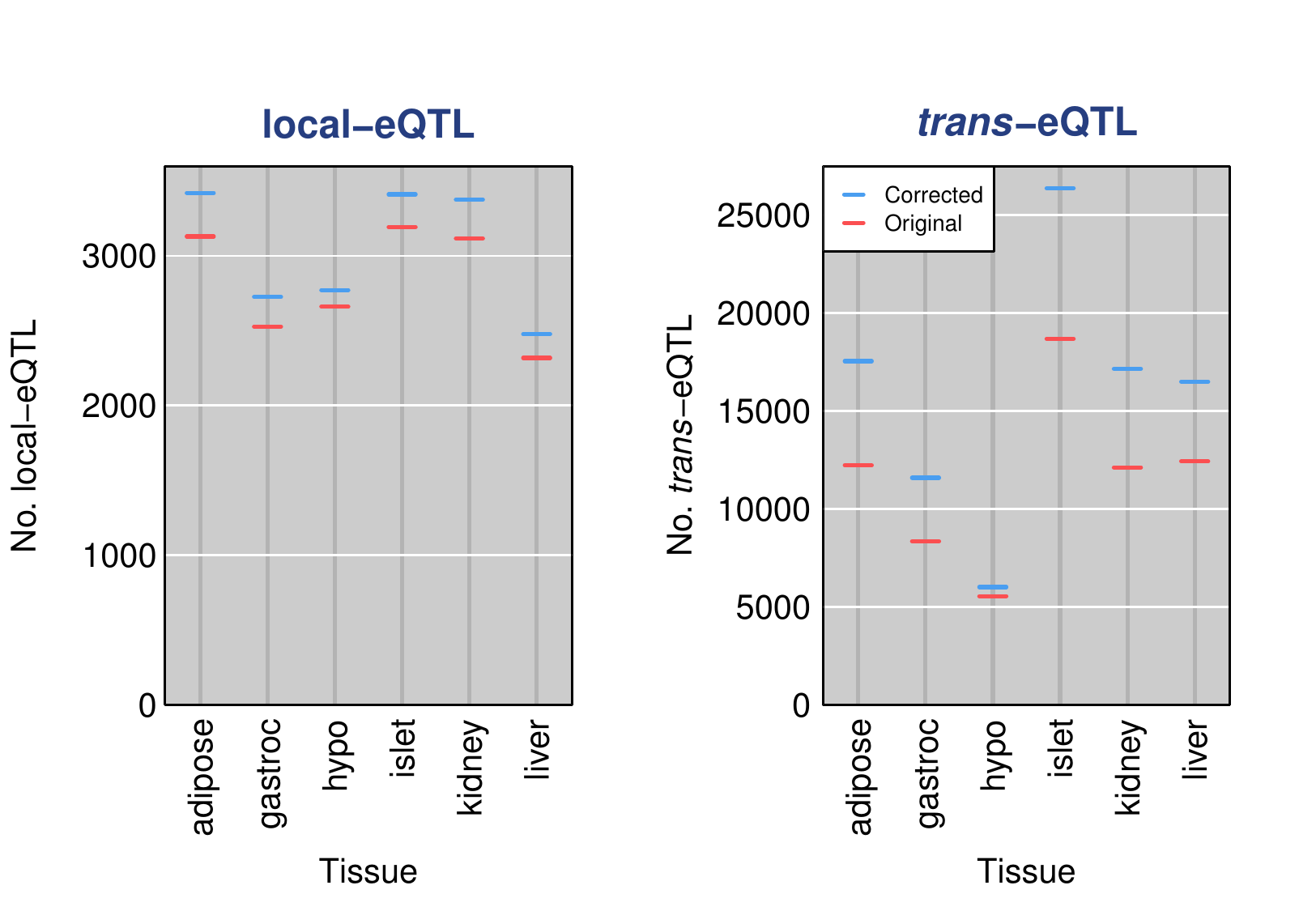}}

\vspace{1cm}

\caption{  Numbers of identified local- and \emph{trans}-eQTL with LOD $\ge$ 5,
  with the original data (red) and after correction of the sample
  mix-ups (blue), across 37,797 array probes with known genomic
  location.  An eQTL was considered local if the 2-LOD support
  interval contained the corresponding probe; otherwise it was
  considered \emph{trans}.
\label{fig:eqtl_counts}}
\end{figure}

Finally the corrected data resulted in a great increase in the numbers
of inferred eQTL in the six tissues (Figure~\ref{fig:eqtl_counts}).
For each array probe with know genomic position, we performed a genome
scan, including sex as an interactive covariate (that is, allowing the
QTL effect to be different in the two sexes).  For each array probe,
we counted the number of chromosomes have a peak LOD score above 5.
Such a peak, on the chromosome containing the probe, was considered a
local-eQTL if the 2-LOD support interval contained the probe location;
other peaks were called \emph{trans}-eQTL.  The inferred number of
local-eQTL increased by 7\% across tissues (with a
somewhat smaller increase in hypo).  The inferred number of
\emph{trans}-eQTL increased by 37\% across tissues
(though only by 8\% in hypo).  The modest
increases in hypo were due in part to the omission of 119 poorly
behaved arrays. The increased numbers of inferred eQTL is also seen
with more stringent thresholds; the numbers of eQTL with LOD $\ge$
10 are shown in Figure~S20.

\clearpage
\centerline{\sffamily \textbf{Discussion}}

In a mouse intercross with over 500 animals and gene expression
microarray data on six tissues, we identified and corrected sample
mix-ups involving 18\% of the DNAs, along with a small number of mix-ups in
each batch of expression arrays.  The QTL mapping results improved
markedly following the correction of mix-ups, but it was perhaps most
surprising just how strong the results were prior to the corrections.

To align the expression arrays, we first identified subsets of genes
with strong between-tissue correlation in expression, and then
considered the correlations between samples across these subsets of
genes.  To align genotypes and expression arrays, we identified
transcripts with strong local-eQTL, formed predictors
of eQTL genotype from expression values, and
calculated the proportion of matches between the observed eQTL
genotypes for a DNA sample and the predicted eQTL genotypes for an
mRNA sample.

This approach applies quite generally: Whenever one has two data
matrices, $X$ and $Y$, whose rows should correspond, one should check that the rows
do in fact correspond.  The simplest approach is to first identify
subsets of associated columns (in which a column of $X$ is associated
with a column of $Y$) and then calculate some measure of similarity
between rows of $X$ and rows of $Y$, across that subset of columns.

Similar approaches have been described by a number of groups.
\citet{Westra2011} considered a number of public
datasets and found an overall rate of 3\% sample mix-ups, with one
dataset \citep{Choy2008} having 23\% mix-ups.
\citet{Schadt2012} showed that, with the tight connection between
genotypes and gene expression phenotypes, external eQTL information
can, in principle, be used to identify individuals participating in a
gene expression study: Genome-wide gene expression is just as
revealing of individual identities as genome-wide genotype data.
\citet{Lynch2012} highlighted issues arising in large tumor studies
and focused particularly on a number of experimental design issues,
such as plate layout.  \citet{Ekstrom2012} considered the
identification of sample mix-ups in genome-wide association studies,
focusing on a small number of phenotypes, such as blood group data,
with strong genotype-phenotype associations.
Also relevant is the forensic bioinformatics work of
\citet{Baggerly2008, Baggerly2009}, particularly
their efforts to correct mix-ups in data files.
Finally, \citet{Jun2012} recently described methods for detecting
mixtures in DNA samples based on genotype or sequencing data,
and there is considerable work on detecting mislabeled microarrays
\citep[e.g.,][]{Zhang2009, Bootkrajang2013}.

There are a number of opportunities for improvement in our approach.
In particular, a number of critical parameters (such as the LOD score
for choosing eQTL, and the number of nearest neighbors and the minimum
vote in the $k$-nearest neighbor
classifier) were chosen in an \emph{ad hoc\/} way.  The choice of such
parameters influences the variation within and the separation between the
self-self and self-nonself distributions of similarity measures, and
thus our ability to identify errors.  In addition, other
classification methods might be used, though the $k$-nearest neighbor
classifier has an important advantage: It works well even in the
presence of mis-classification error in the ``training'' data.

Perhaps the most important lesson from this work is the value of
investigating aberrations.  One should follow up any observed
inconsistencies in data, to identify the source.  In particular,
one should not rely solely on LOD scores or other summary statistics,
but also inspect plots of genotype versus phenotype, such as that
in Figure~\ref{fig:gve}.

Of course, there are many possible errors that we couldn't see by
these approaches.  For example, all of the tissues (including the DNA)
for a pair of animals might be swapped, or there may be mix-ups within
the clinical phenotypes (such as plasma insulin levels).  And some
mix-ups are detectable but not correctable.

We have not identified any between-tissue mix-ups in the expression
data, but such errors are possible.  For that type of error, it may be
useful to consider the gene expression bar code developed by
\citet{Zilliox2007}.

The correction of inferred sample mix-ups, as we have done, may
introduce bias towards larger estimated eQTL effects.  We believe
that, in the current study, there is little risk of such bias, as the
data provide strong evidence for specific sample labels.  If the
correction of sample mix-ups were accompanied by a higher level of
uncertainty, one might consider omitting samples rather than assigning
the inferred labels, though such an approach could also incur some
bias.

Finally, one might ask, following these findings: What is an
acceptable error rate in a research study?  And what laboratory
procedures should be instituted to avoid such errors? There exist
procedures to help protect against errors, both for genotypes
\citep[e.g.,][]{Huijsmans2007a, Huijsmans2007b} and for microarrays
\citep{Grant2003, Imbeaud2005, Walter2010}, but they are not always
put into practice.  However, as the current study indicates, with
expression genetic data, one can accommodate a high rate of errors
provided that one applies appropriate procedures to detect and correct
such errors.

\clearpage
\centerline{\sffamily \textbf{Acknowledgments}}

The authors thank Angie Oler, Mary Rabaglia, Kathryn Schueler, and
Donald Stapleton for their work on the underlying project, and Amit
Kulkarni for providing annotation information for the expression
microarrays.
This work was supported in part by
National Institutes of Health grants GM074244 (to K.W.B.) and
DK066369 (to A.D.A).

\clearpage
\bibliographystyle{genetics}
\renewcommand*{\refname}{\centerline{\normalsize\sffamily \textbf{Literature Cited}}}
\bibliography{samplemixups}

\clearpage

\renewcommand{\thefigure}{\textbf{S\arabic{figure}}}
\renewcommand{\figurename}{\textbf{Figure}}

\renewcommand{\thetable}{\textbf{S\arabic{table}}}
\renewcommand{\tablename}{\textbf{Table}}

\setcounter{figure}{0}

\vspace*{8mm}
\begin{center}

\textbf{\Large Identification and correction of sample mix-ups \\[14pt]
in expression genetic data: A case study}

\bigskip \bigskip \bigskip \bigskip

\textbf{\Large SUPPLEMENT}

\bigskip \bigskip
\bigskip \bigskip

{\large Karl W. Broman$^*$,
Mark P. Keller$^{\dagger}$,
Aimee Teo Broman$^{*}$, \\[8pt]
Christina Kendziorski$^{*}$,
Brian S Yandell$^{\ddagger, \S}$,
\'Saunak Sen$^{**}$,
Alan D. Attie$^{\dagger}$}

\bigskip \bigskip

$^{*}$Department of Biostatistics and Medical Informatics,
$^{\dagger}$Department of Biochemistry,
$^{\ddagger}$Department of Statistics,
and $^{\S}$Department of Horticulture,
University of Wisconsin--Madison, Madison, Wisconsin
53706, and
$^{**}$Department of Epidemiology and Biostatistics, University
of California, San Francisco, California 94107
\end{center}

\clearpage

\begin{figure}[p]
\centerline{\includegraphics{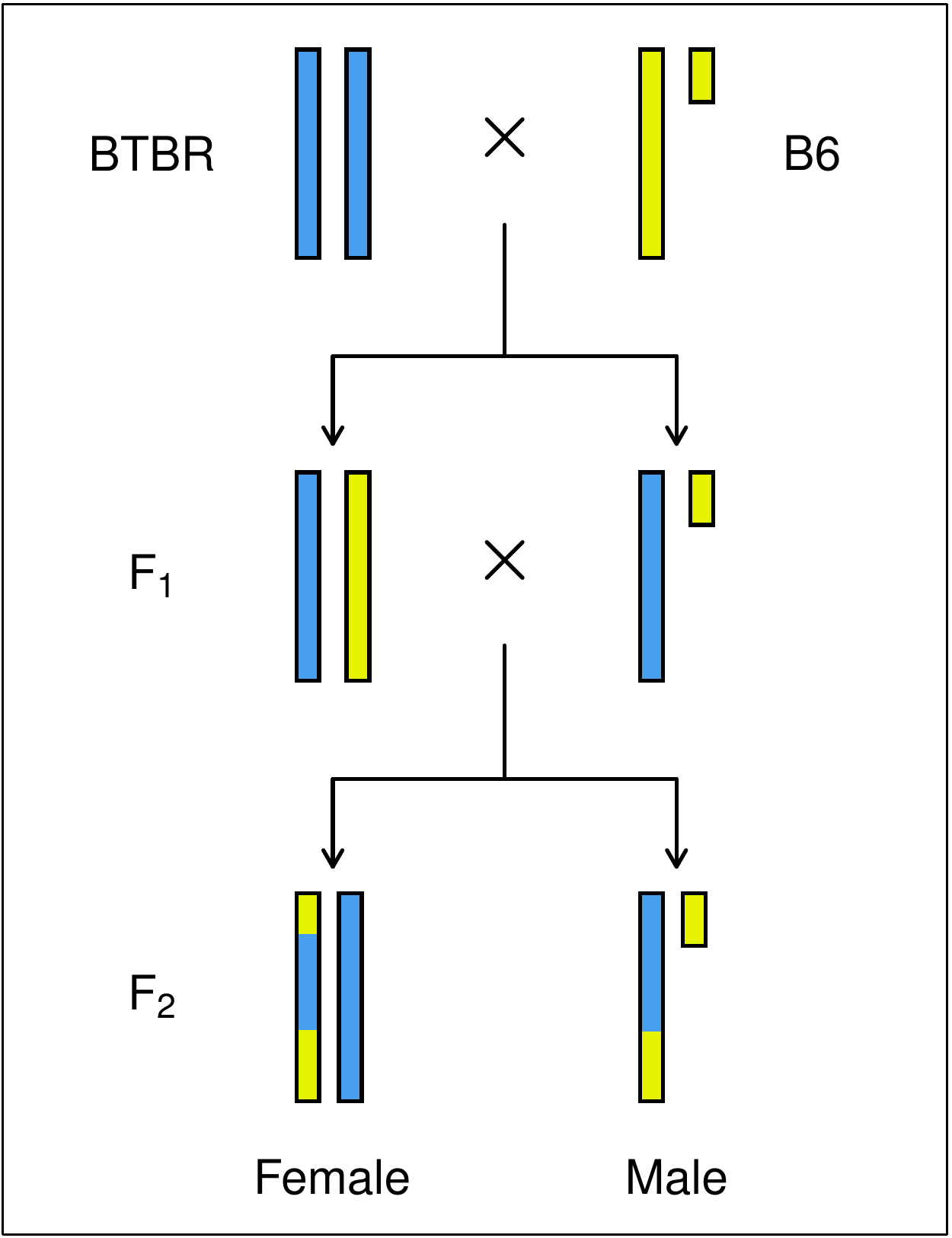}}

\caption{The behavior of the X chromosome
in the intercross (BTBR $\times$ B6) $\times$ (BTBR $\times$ B6).  In
the F$_2$ generation, females are homozygous BTBR or heterozygous,
while males are hemizygous BTBR or B6.  The small bar is the Y
chromosome.
}
\end{figure}

\clearpage

\begin{figure}[p]
\centerline{\includegraphics[angle=270,width=\textwidth]{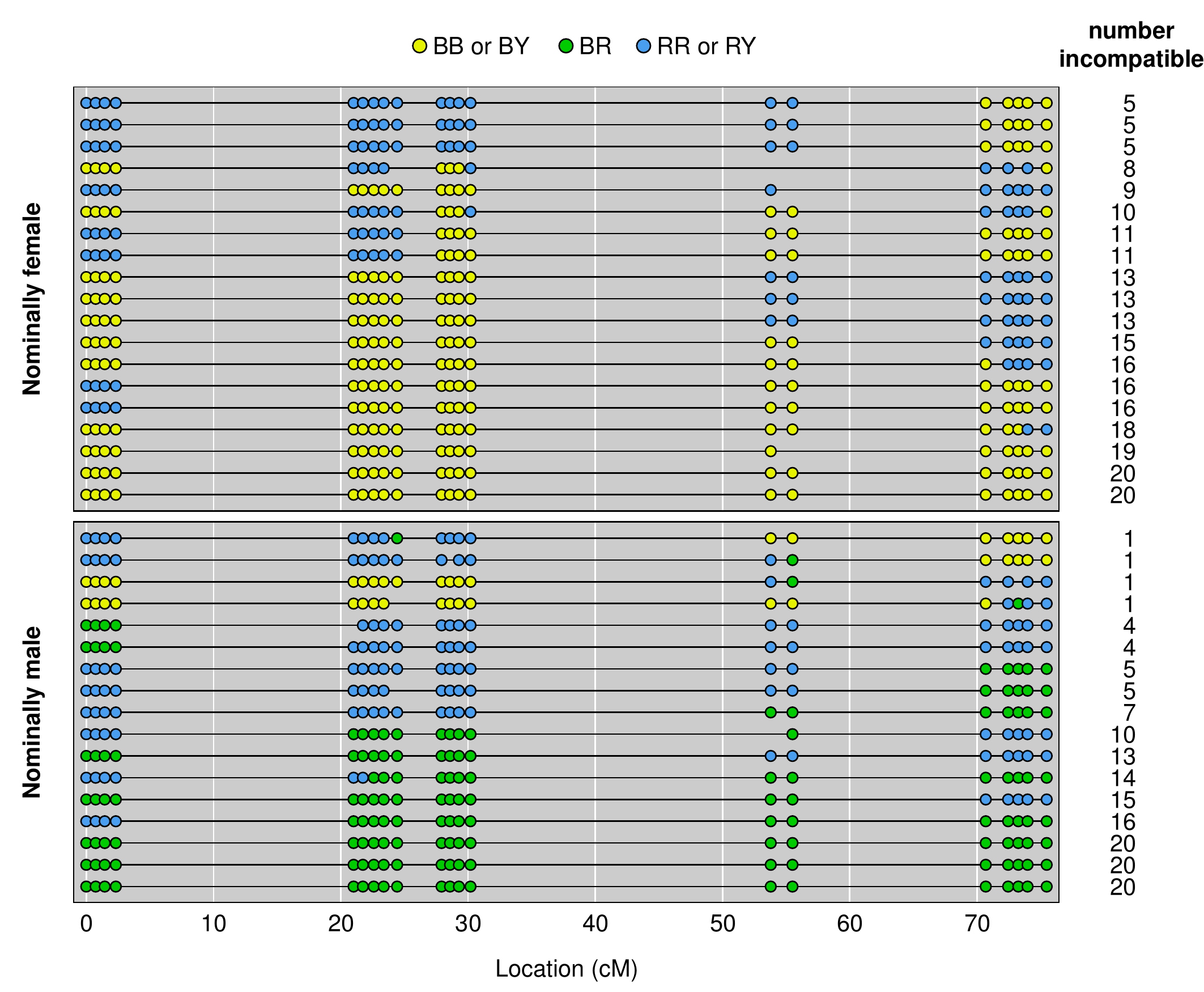}}

\caption{X chromosome genotypes for
19~female mice and 17~male mice with genotypes that are incompatible
with their sex.  Females should be homozygous BTBR (RR, blue) or
heterozygous (green).  Males should be hemizygous B6 (BY, yellow) or
hemizygous BTBR (RY, blue).  The top four males have a single
incompatibility that could reasonably be a genotyping error.
}
\end{figure}

\clearpage

\begin{figure}[p]
\centerline{\includegraphics[width=4.5in]{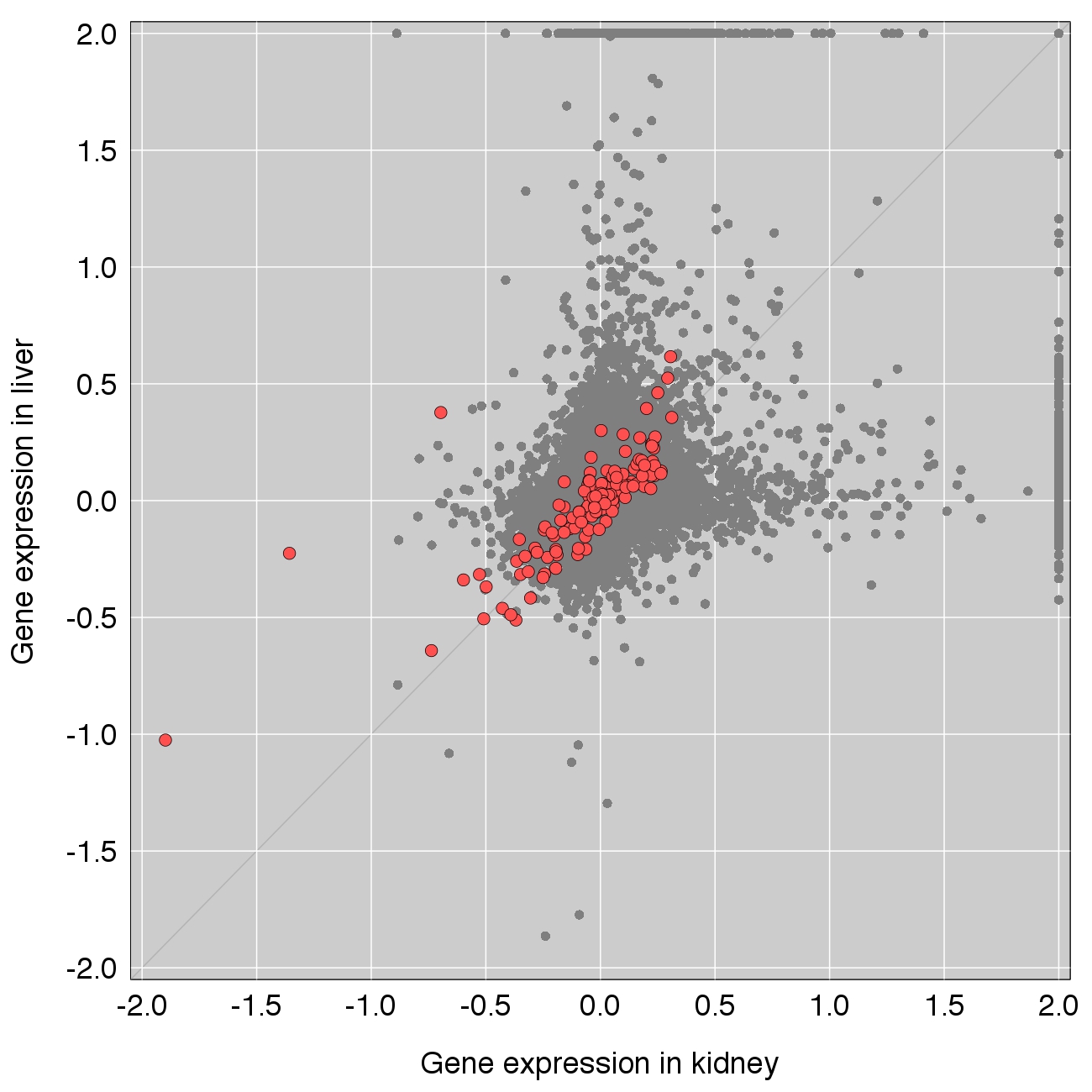}}

\caption{Example scatterplot of gene expression in
liver versus kidney for a single individual (Mouse3567).  Gray points
are all probes on the array; red points are the 155 probes with correlation
across mice $>$ 0.75 between liver and kidney.
}
\end{figure}

\clearpage

\begin{figure}[p]
\centerline{\includegraphics{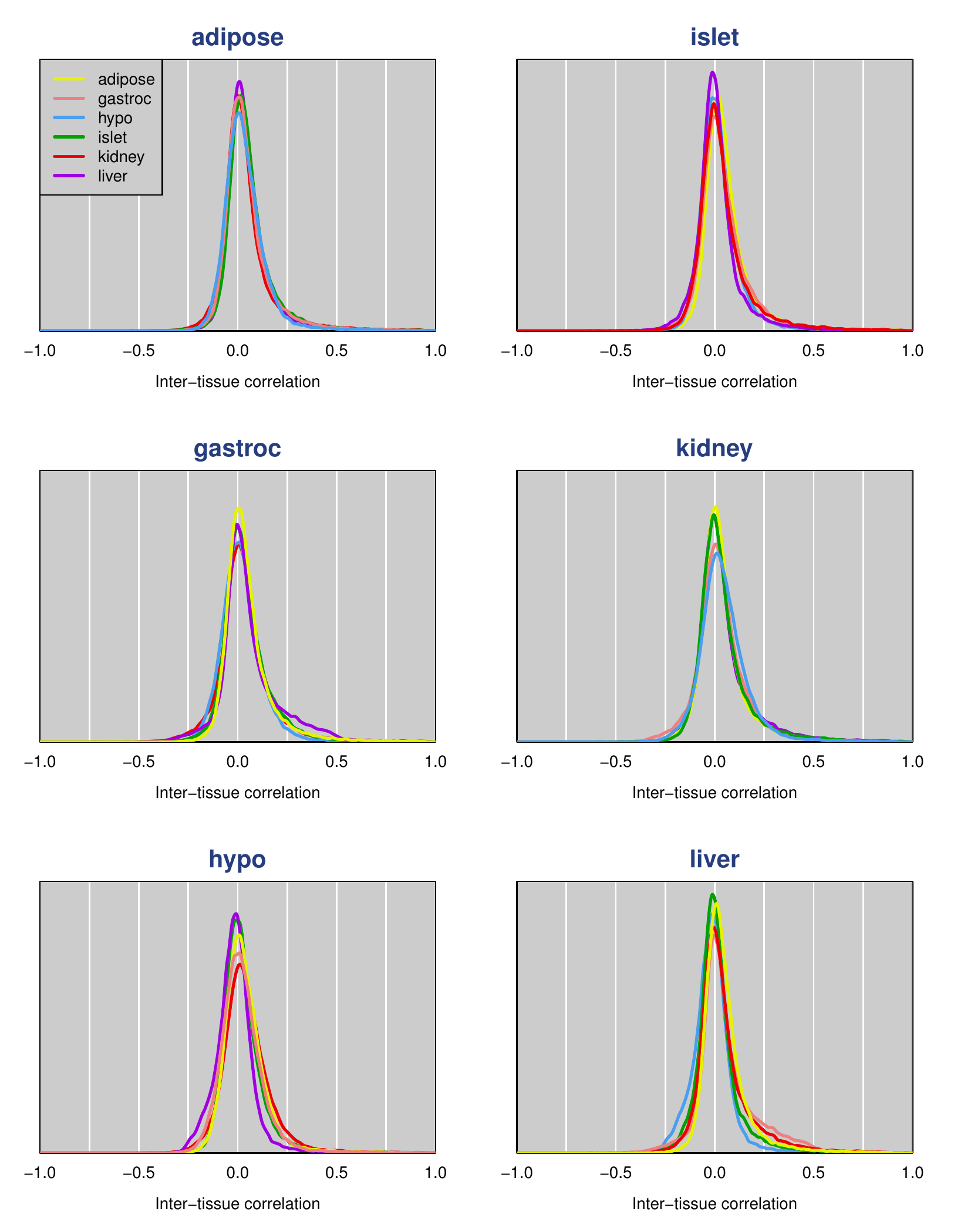}}

\caption{Density estimates of the
between-tissue correlations for all probes on the expression arrays.
In each panel, the distributions for the five pairs of tissues,
including a given tissue, are displayed.
}
\end{figure}

\clearpage

\begin{figure}[p]
\centerline{\includegraphics{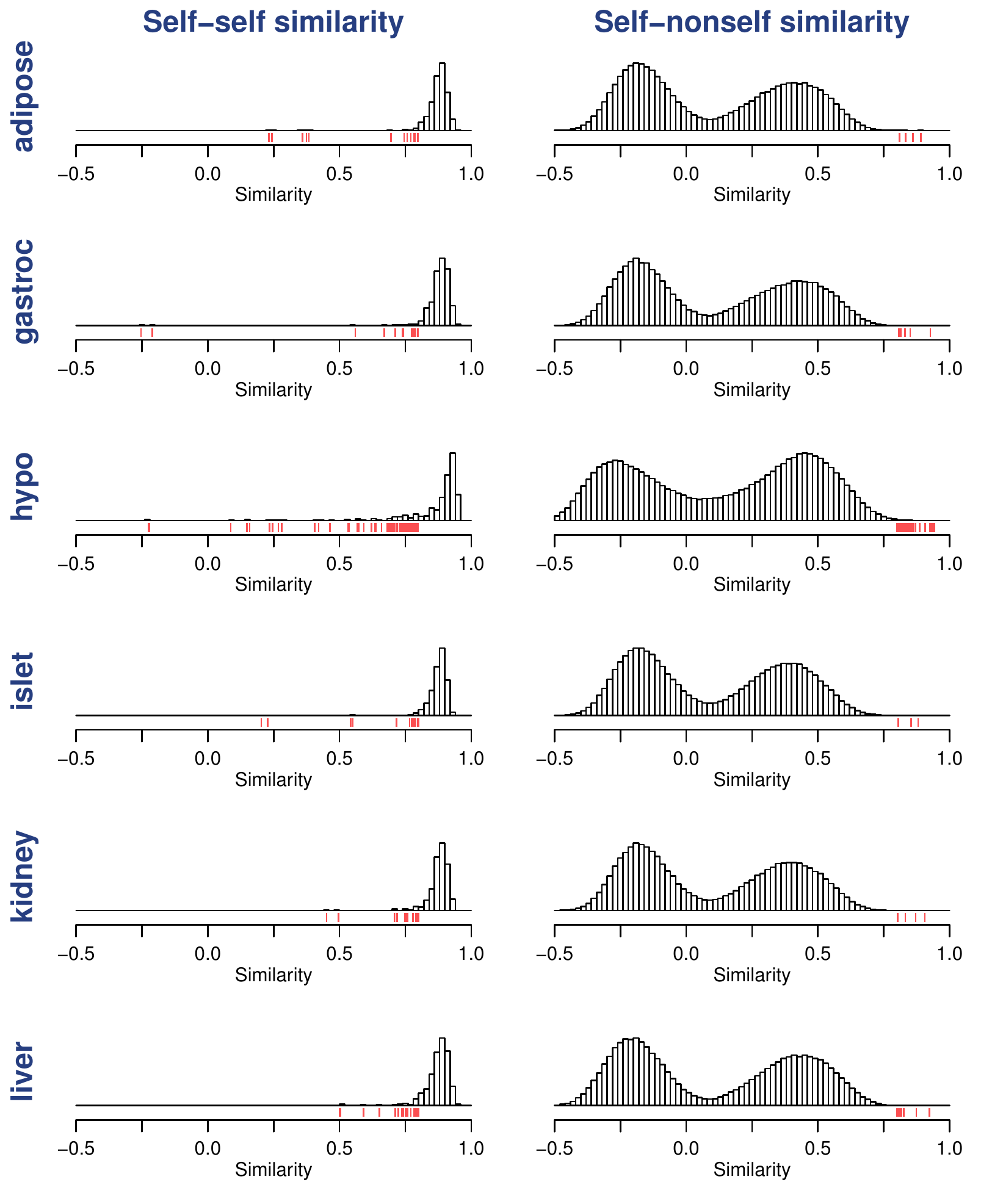}}

\caption{Histograms of similarity measures for the expression arrays
  for each tissue, versus all other tissues combined.  The panels on
  the left include self-self similarities (along the diagonal of the
  similarity matrices); the panels on the right include all
  self-nonself similarities (the off-diagonal elements of the
  similarity matrices). Self-self values $<$ 0.8 and self-nonself values
  $>$ 0.8 are highlighted with red tick marks.  The two modes in the
  self-nonself distributions are for opposite-sex and same-sex pairs.
}
\end{figure}

\clearpage

\begin{figure}[p]
\centerline{\includegraphics{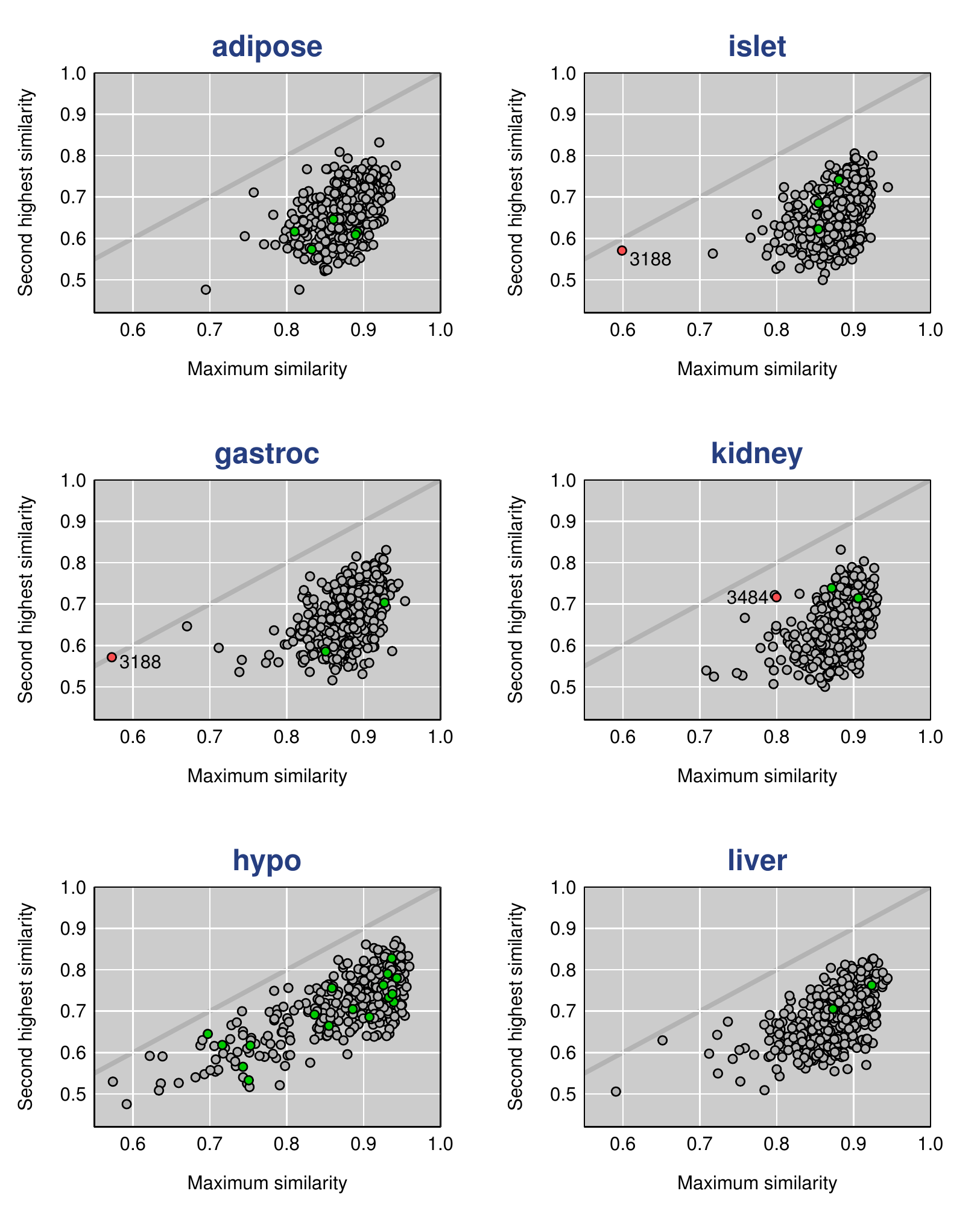}}

\caption{Second highest similarity (median
correlation across tissue pairs) versus maximum similarity for the
expression arrays for each tissue.  The diagonal gray line corresponds
to equality.  Green points correspond to arrays inferred to be sample
mix-ups.  Gray points correspond to arrays for which the self
similarity is maximal.  Red points correspond to special cases, as in
Figure~1 (see the text).
}
\end{figure}

\clearpage

\begin{figure}[p]
\centerline{\includegraphics{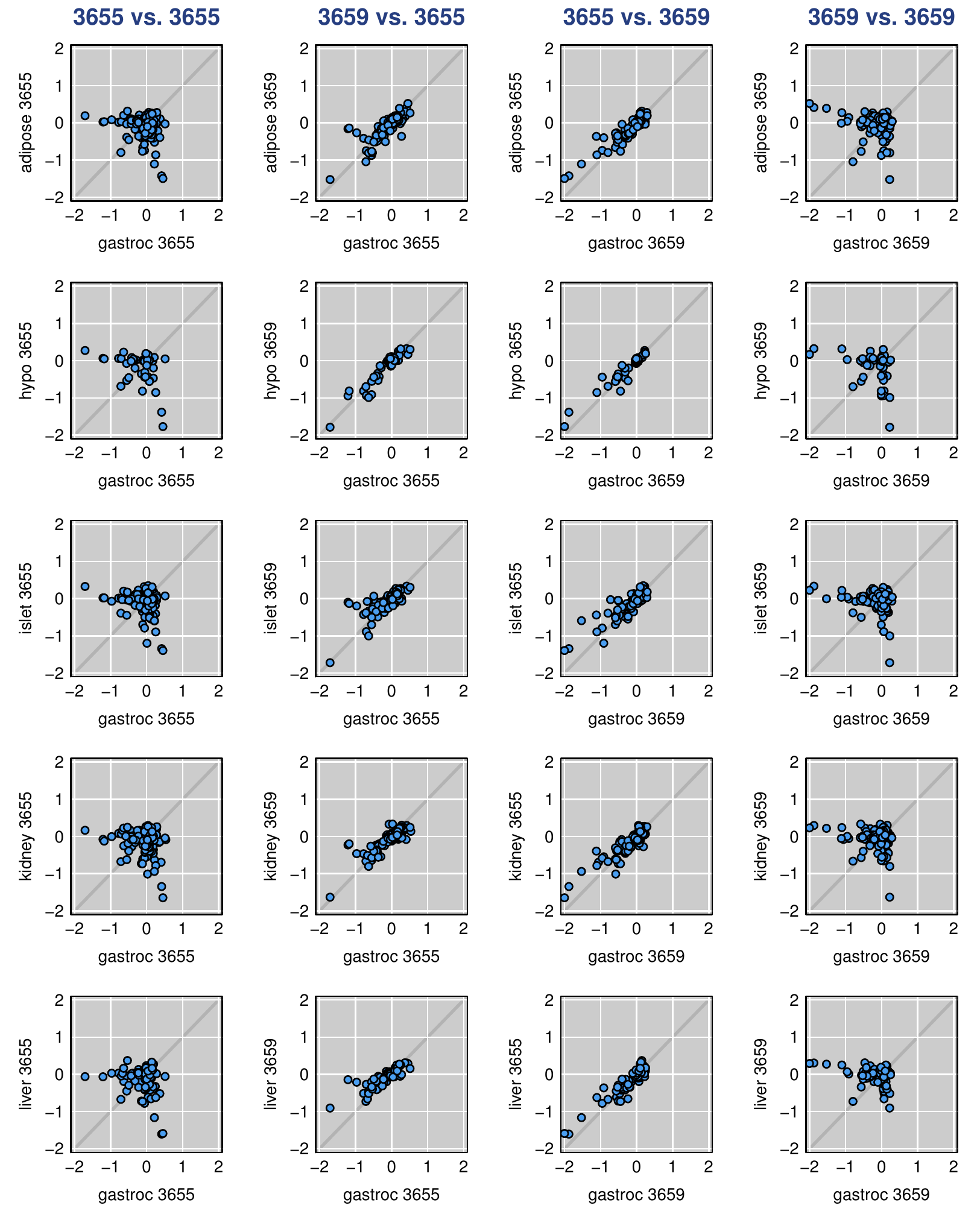}}

\caption{Scatterplots for expression in
pairs of tissues for an inferred sample swap, between Mouse3655 and
Mouse3659 in gastroc.
}
\end{figure}

\clearpage

\begin{figure}[p]
\centerline{\includegraphics{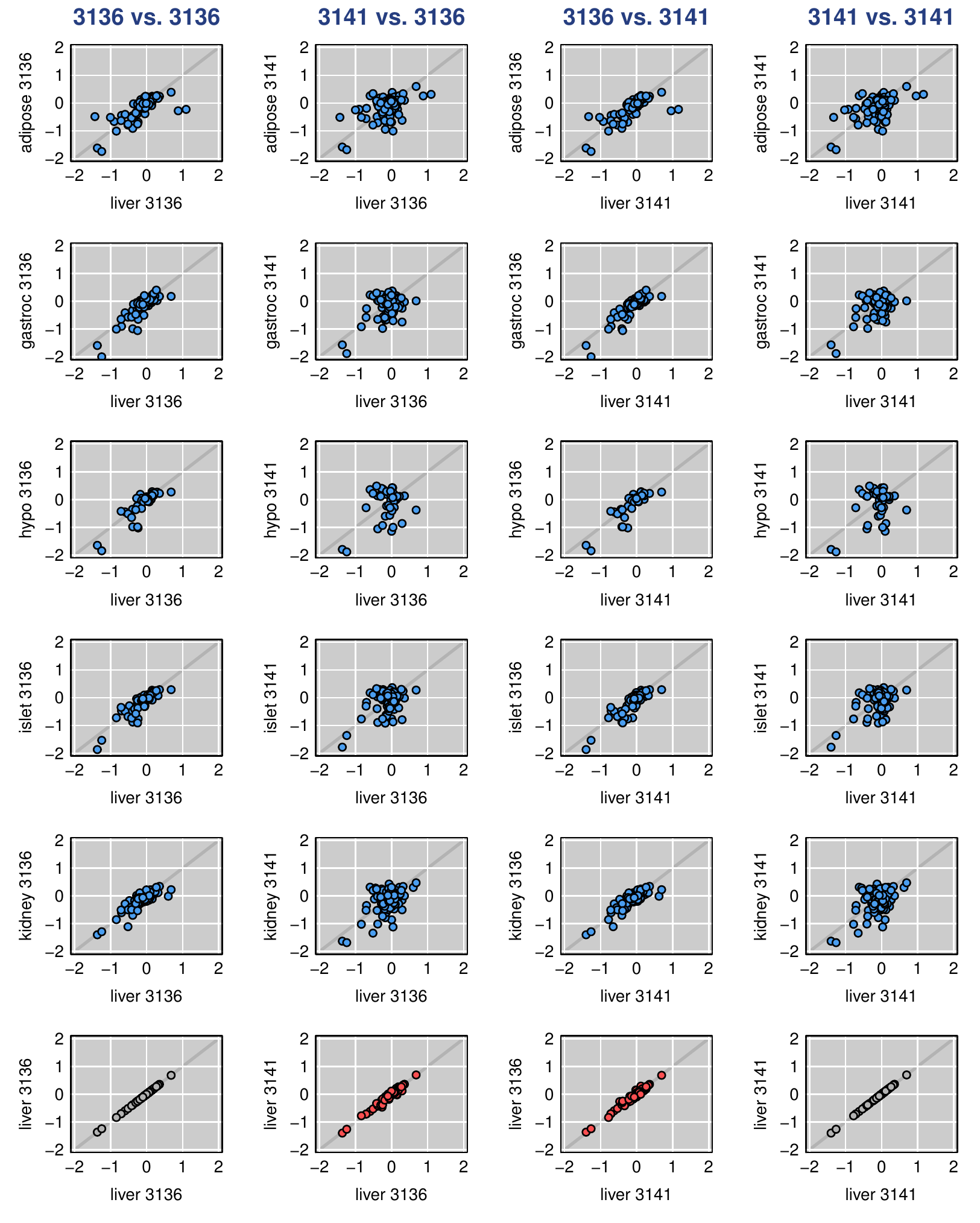}}

\caption{Scatterplots for expression in
pairs of tissues for an inferred sample duplicate, with Mouse3136 in
liver also arrayed as Mouse3141 liver.  In the bottom row, the panels
with gray points are identical data, and the panels with red points
are the unintended duplicates.
}
\end{figure}

\clearpage

\begin{figure}[p]
\centerline{\includegraphics{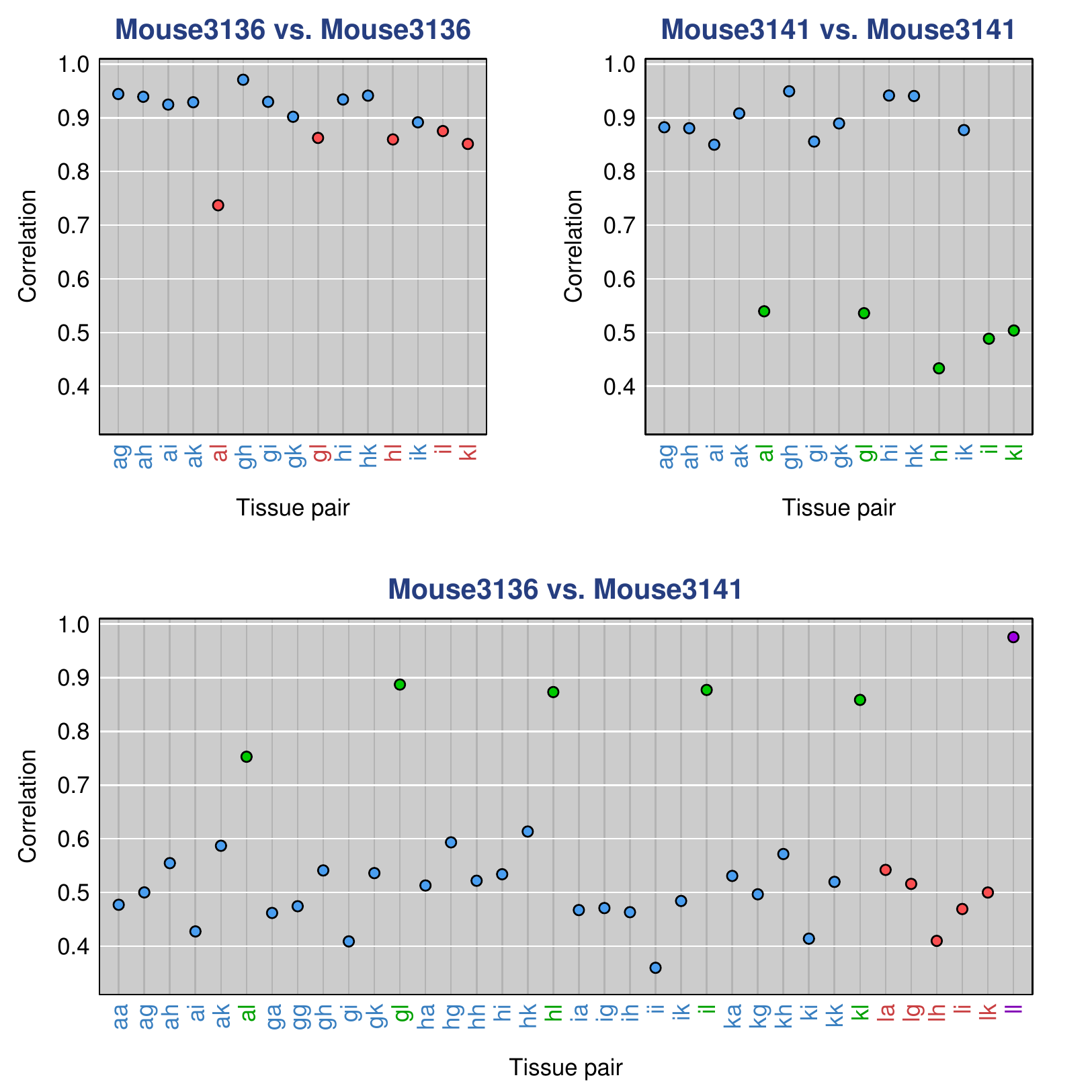}}

\caption{Between-tissue correlations for
pairs of tissues for an inferred sample duplicate, with Mouse3141 in
liver really being a duplicate of Mouse3136 in liver.
Correlations are calculated
using tissue-pair-specific probes that show between-tissue
correlation, across all mice, of $>$ 0.75.  Tissue pairs are
abbreviated by the first letter of the tissues' names.  Red points
involve Mouse3136 liver, green points involve Mouse3141 liver, and the
purple point involves both.
}
\end{figure}

\clearpage

\begin{figure}[p]
\centerline{\includegraphics{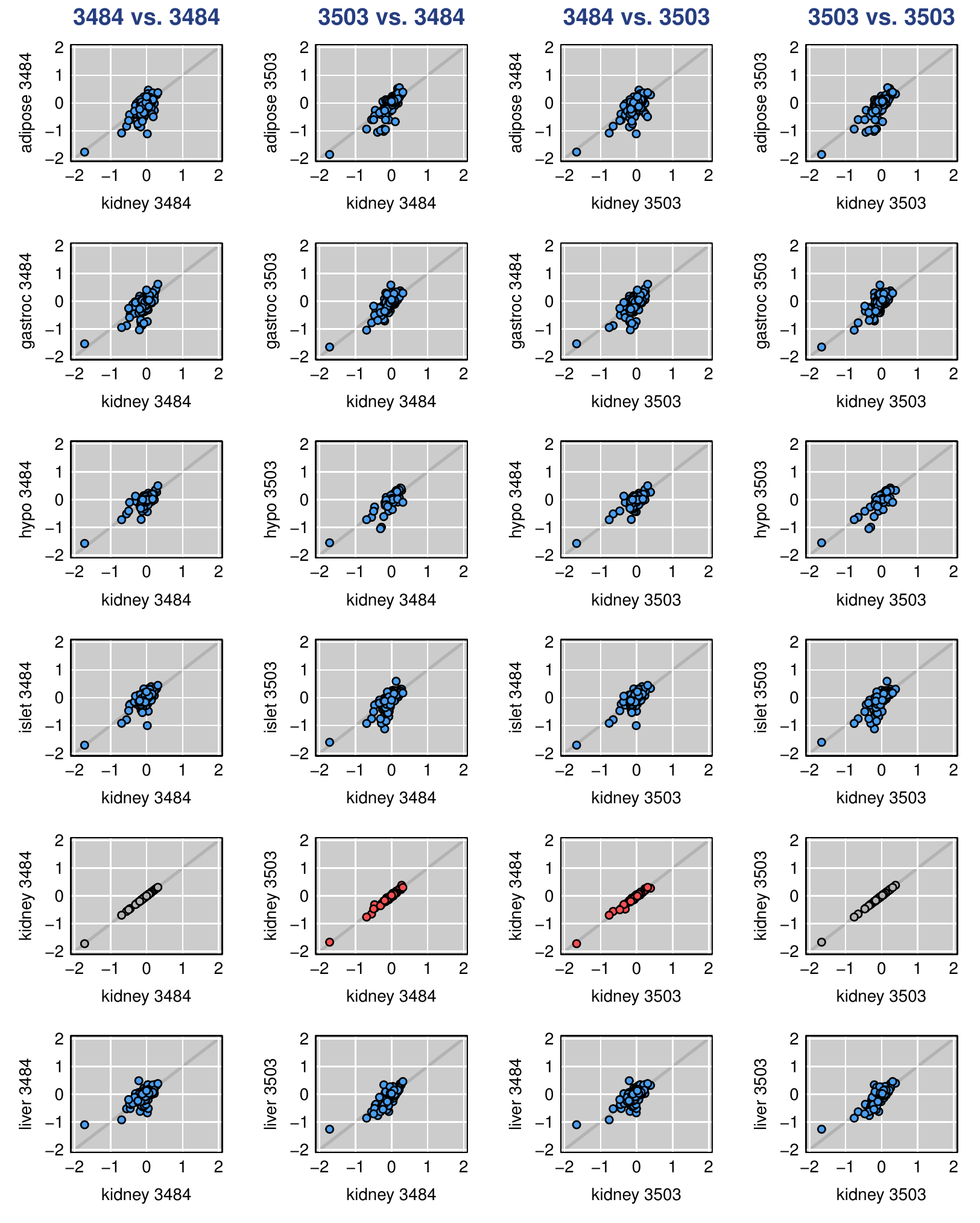}}

\caption{Scatterplots for expression in
pairs of tissues for a potential sample mixture, of Mouse3484 and
Mouse3503 in kidney.  In the second from the bottom row, the panels
with gray points are identical data, and the panels with red points
are the unintended duplicates.
}
\end{figure}

\clearpage

\begin{figure}[p]
\centerline{\includegraphics{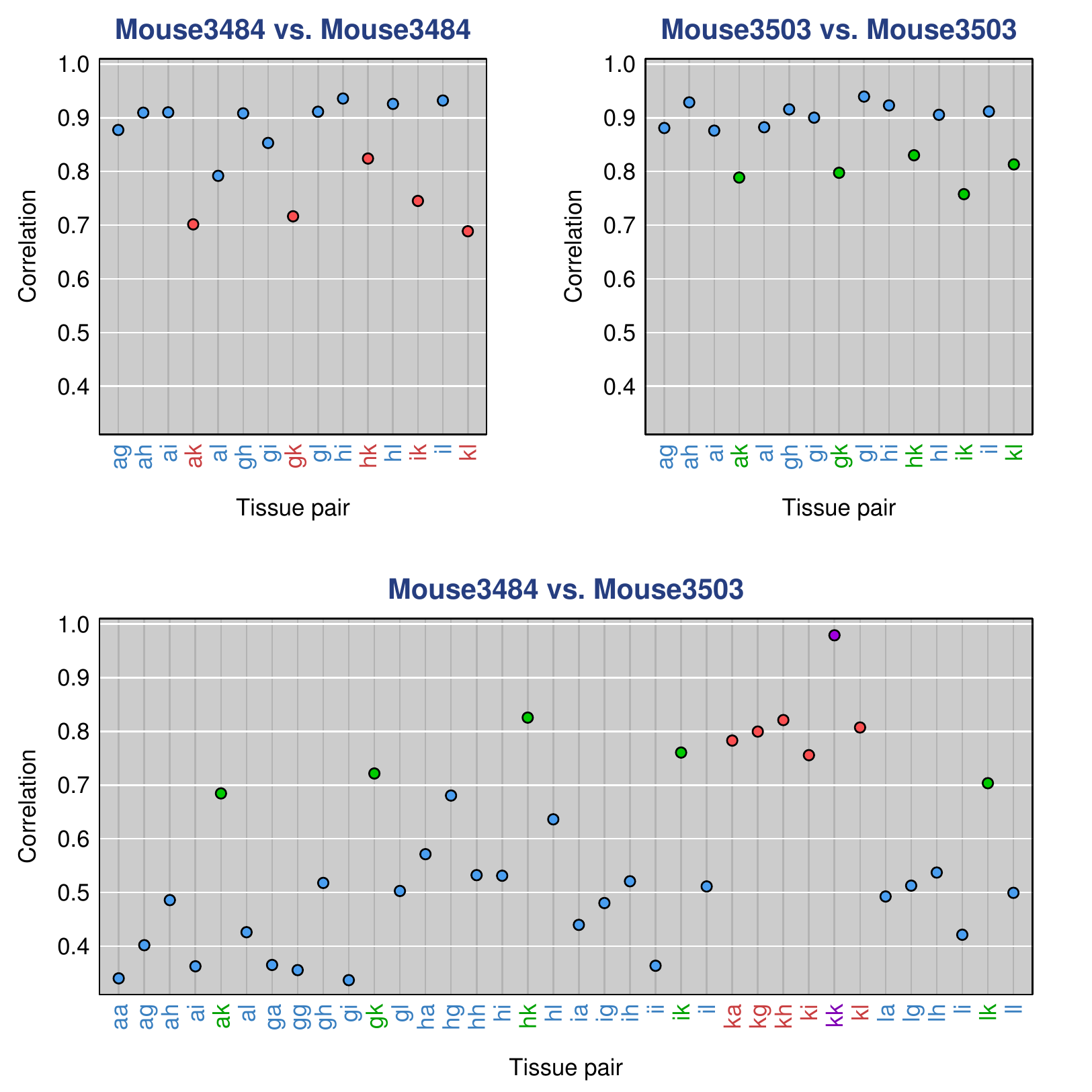}}

\caption{Between-tissue correlations for
pairs of tissues for a potential sample mixture, of Mouse3484 and
Mouse3503 in kidney.  Correlations are calculated using
tissue-pair-specific probes that show between-tissue correlation,
across all mice, of $>$ 0.75.  Tissue pairs are abbreviated by the
first letter of the tissues' names.  Red points
involve Mouse3484 kidney, green points involve Mouse3503 kidney, and the
purple point involves both.
}
\end{figure}

\clearpage

\begin{figure}[p]
\centerline{\includegraphics[height=8.0in]{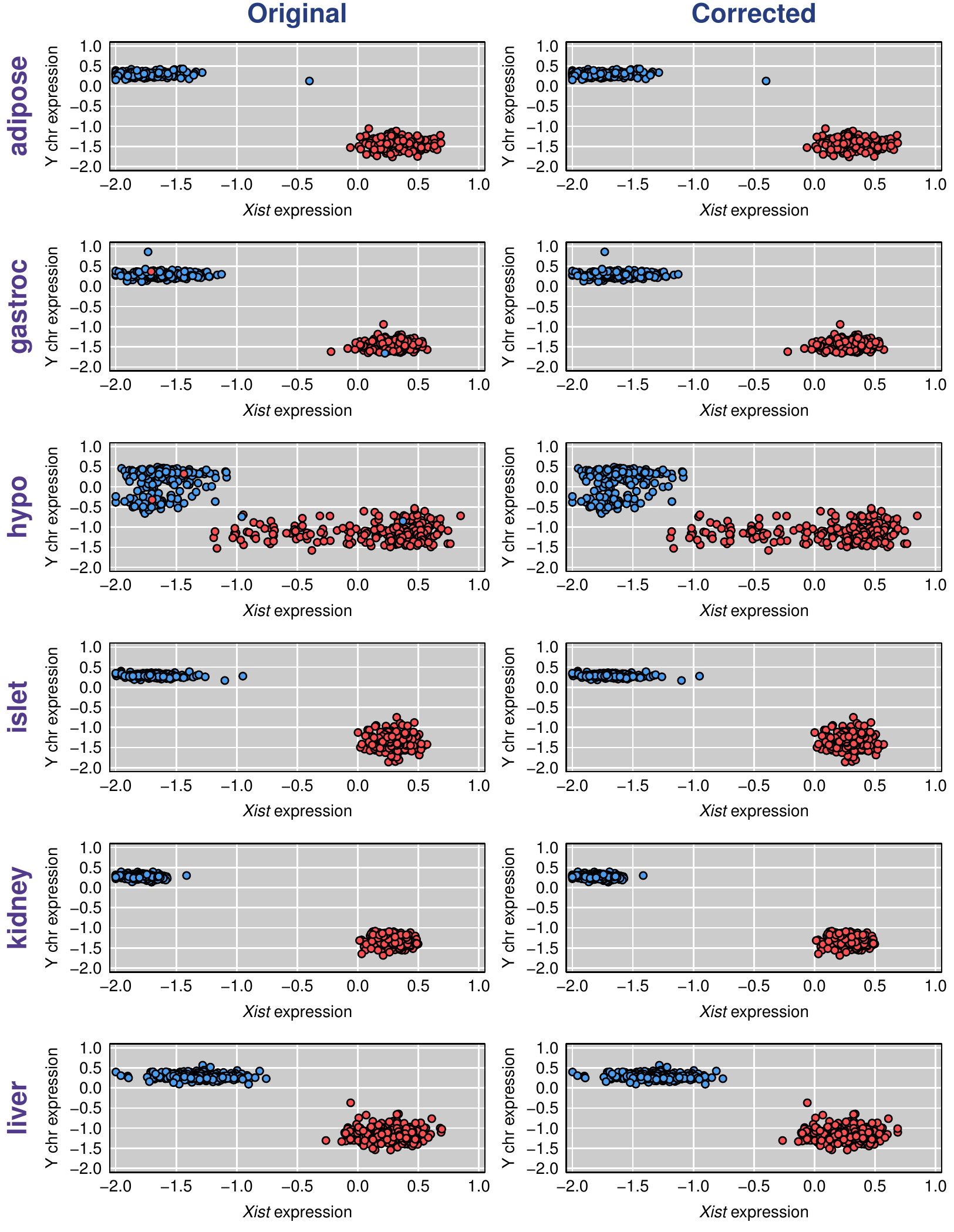}}

\caption{Scatterplots of the average
expression for four Y chromosome genes versus expression of the \emph{Xist\/}
gene in each tissue, before and after correction of sample mix-ups.
Females are in red; males are in blue.  The unusual pattern in
hypothalamus is due to a batch of 120 poorly behaved arrays.
}
\end{figure}

\clearpage

\begin{figure}[p]
\centerline{\includegraphics{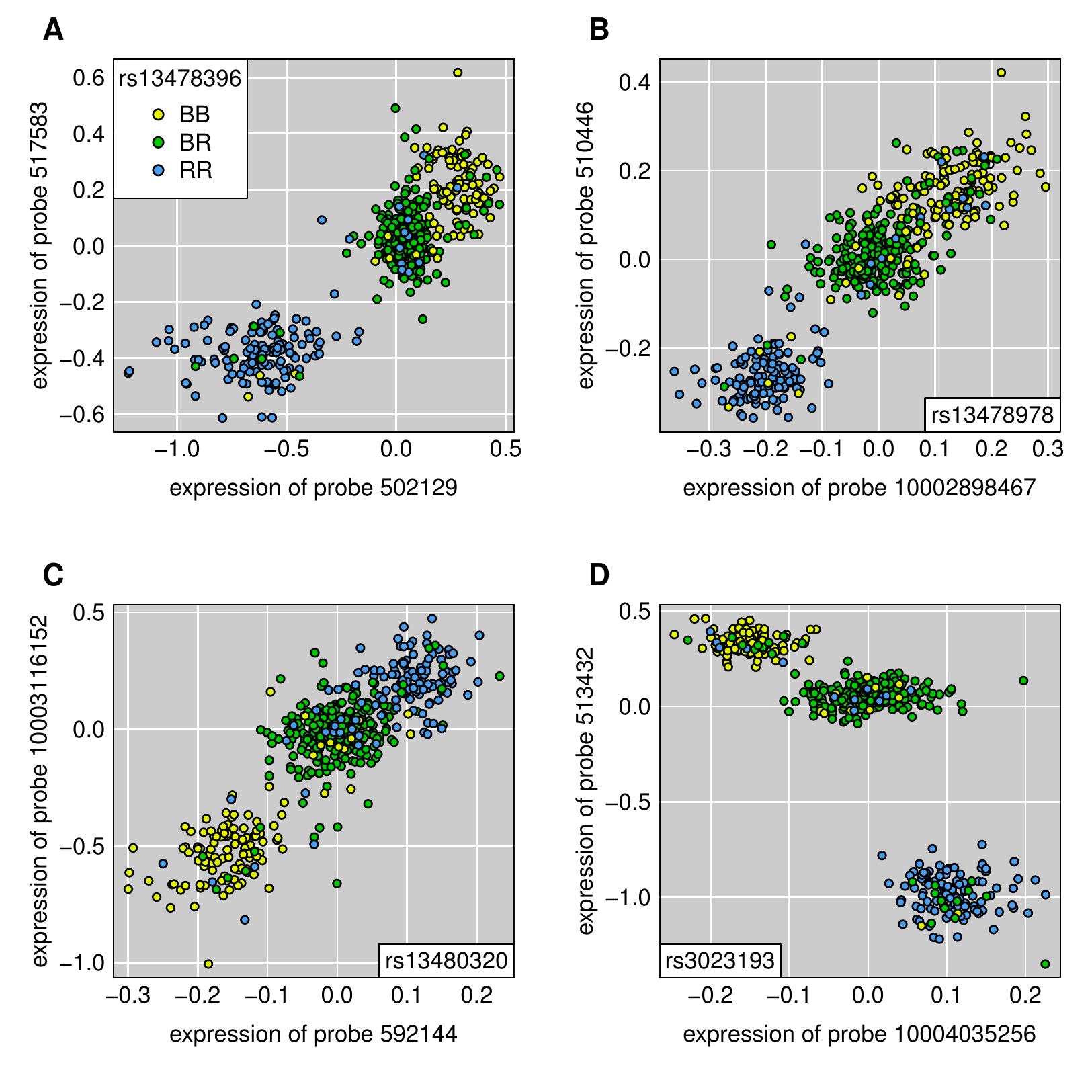}}

\caption{Example scatterplots of islet expression
for pairs of probes at the same genomic location.
}
\end{figure}

\clearpage

\begin{figure}[p]
\centerline{\includegraphics[height=8.0in]{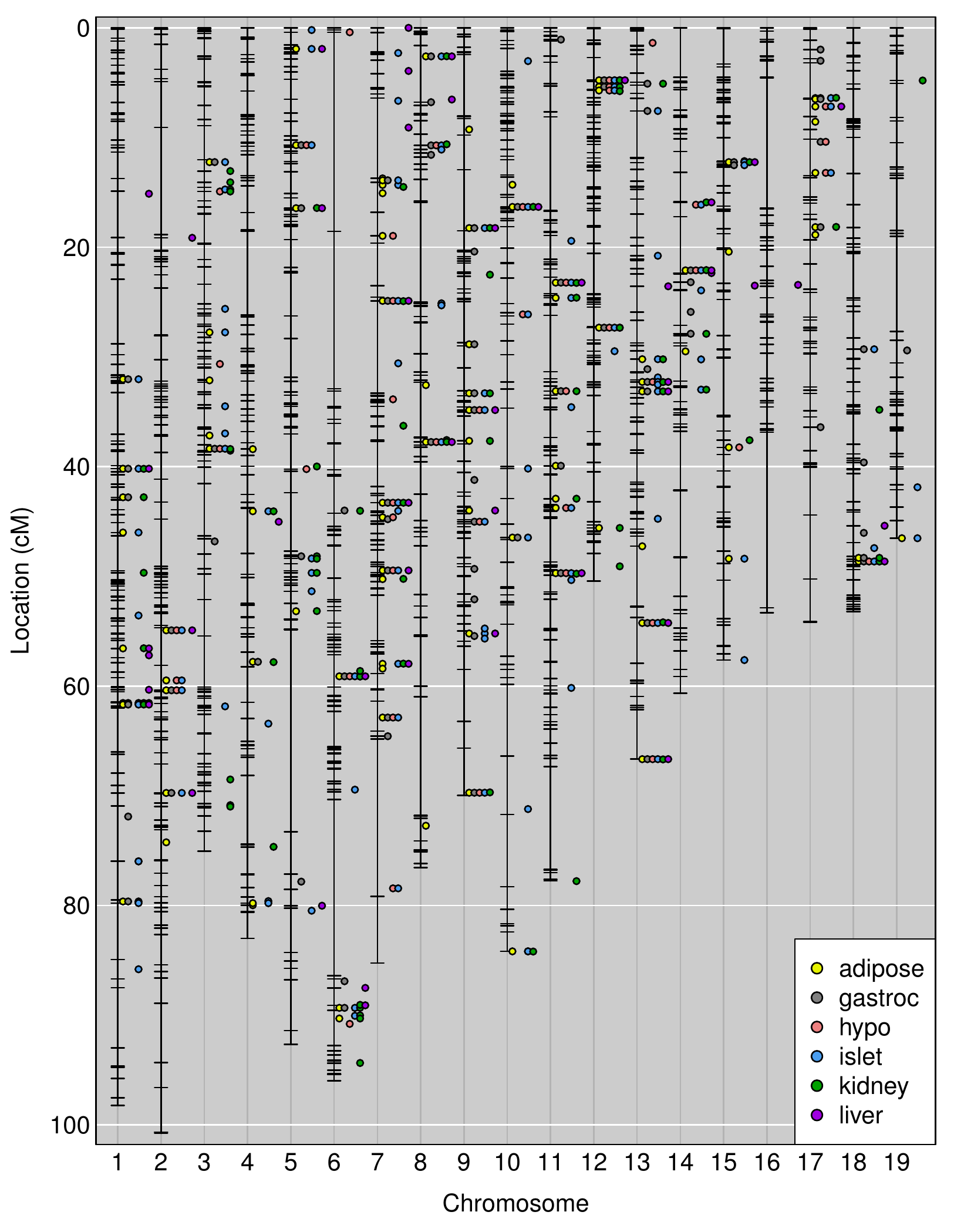}}

\caption{Positions of local eQTL used for
the aligning the expression arrays and genotype data.  Marker
locations are indicated by horizontal line segments on the genetic
map.  The points to the right of each chromosome indicate the eQTL
locations, with different colors for different tissues.
}
\end{figure}

\clearpage

\begin{figure}[p]
\centerline{\includegraphics{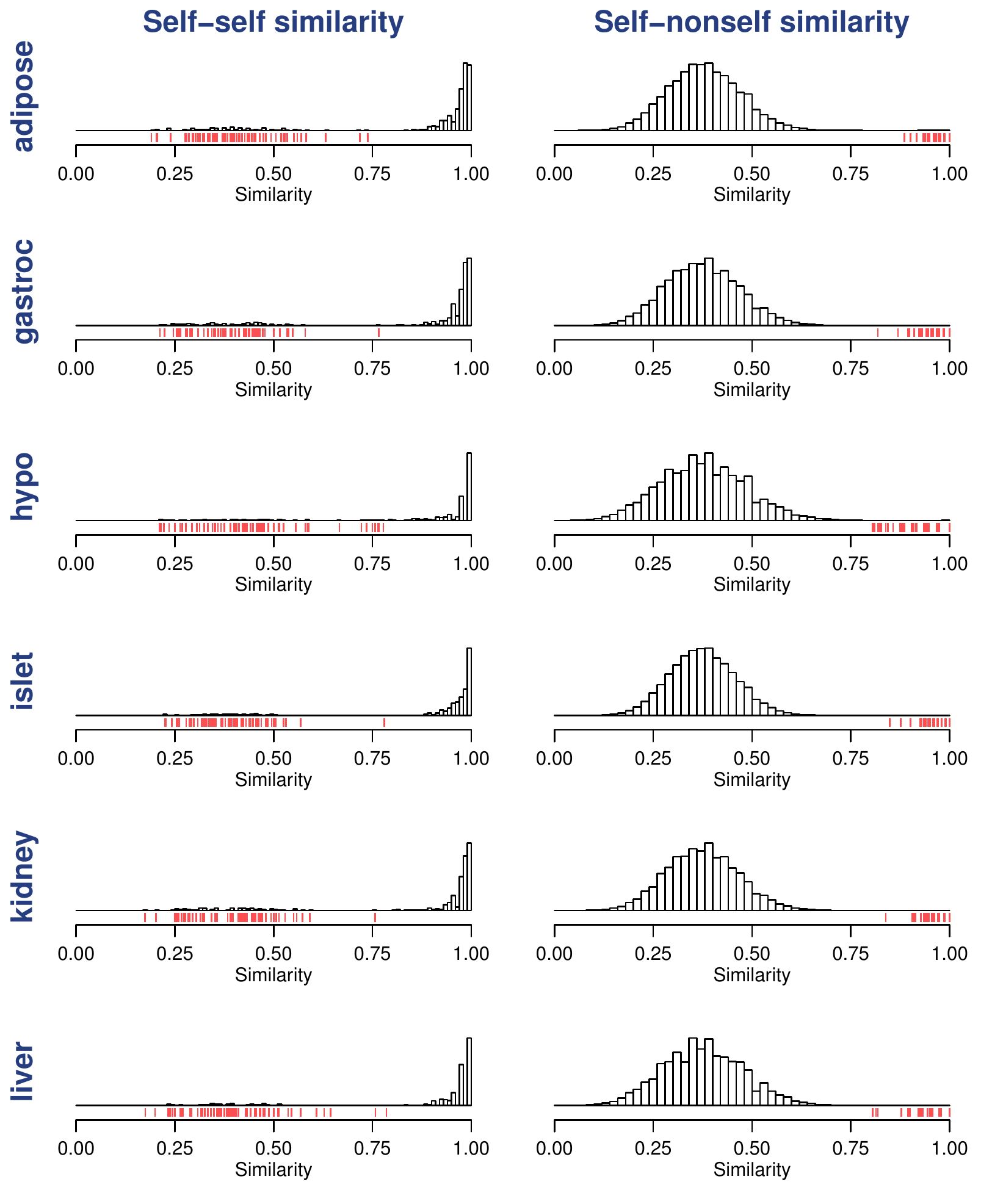}}

\caption{  Histograms of similarities between the genotypes and the
  expression arrays (the proportion of matches between observed and
  inferred eQTL genotypes)
  for each tissue.  The panels on
  the left include self-self similarities (along the diagonal of the
  similarity matrices); the panels on the right include all
  self-nonself similarities (the off-diagonal elements of the
  similarity matrices). Self-self values $<$ 0.8 and self-nonself values
  $>$ 0.8 are highlighted with red tick marks.
}
\end{figure}

\begin{figure}[p]
\centerline{\includegraphics[height=7.65in]{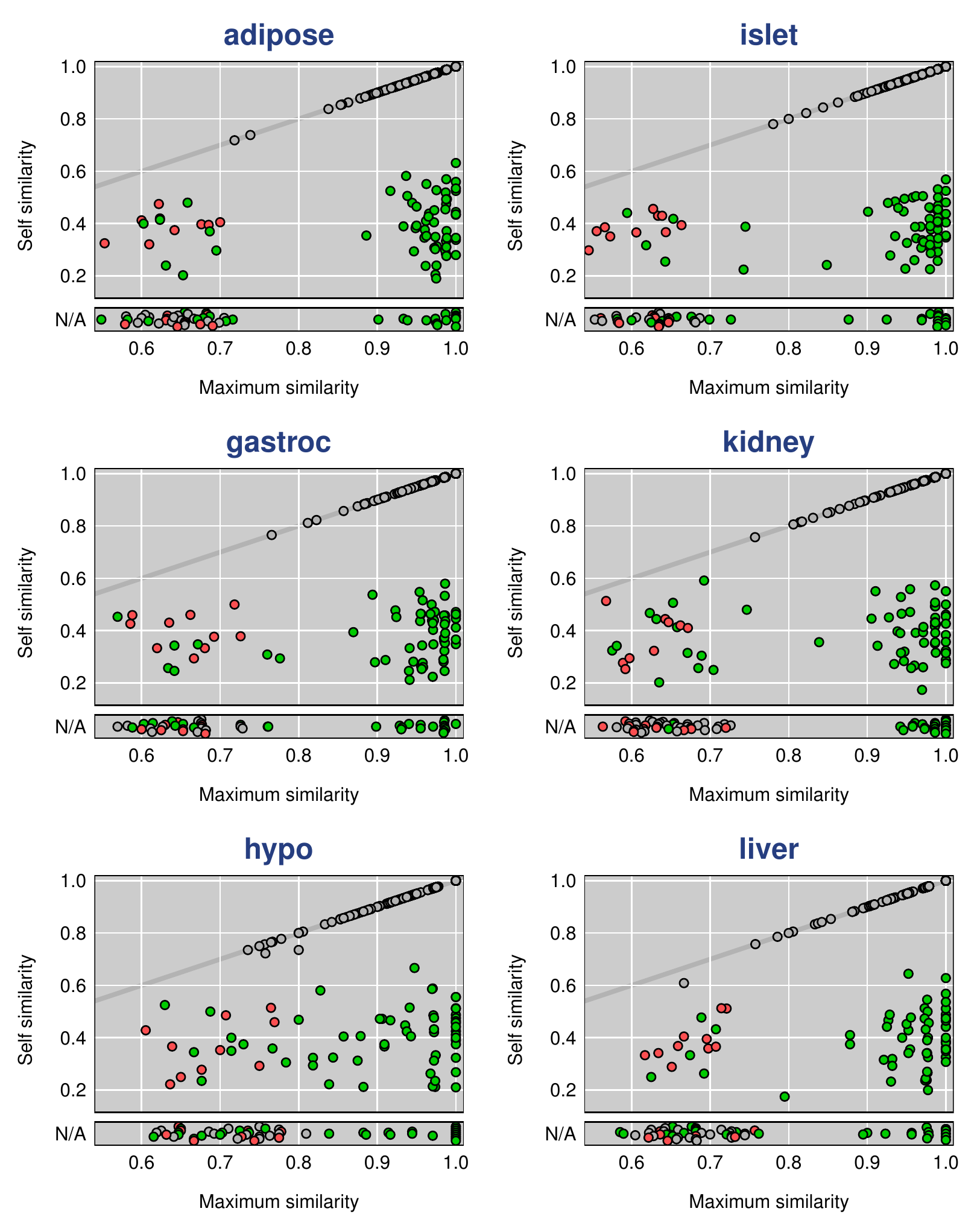}}

\caption{Self similarity (proportion matches between observed and
  inferred eQTL genotypes, considering each tissue separately) versus
  maximum similarity for the DNA samples.  The diagonal gray line
  corresponds to equality.  Samples with missing self similarity (at
  top) did not have an expression assay performed for that tissue.
  Points are colored based on the inferred status of the corresponding
  samples based on the combined information from all tissues.  Gray
  points correspond to DNA samples that were correctly labeled.  Green
  points correspond to sample mix-ups that are fixable (the correct
  label can be determined).  Red points comprise both samples mix-ups
  that cannot be corrected as well as samples that may be correct but
  cannot be verified as no expression data is available.
}
\end{figure}

\clearpage

\begin{figure}[p]
\centerline{\includegraphics[width=4.5in]{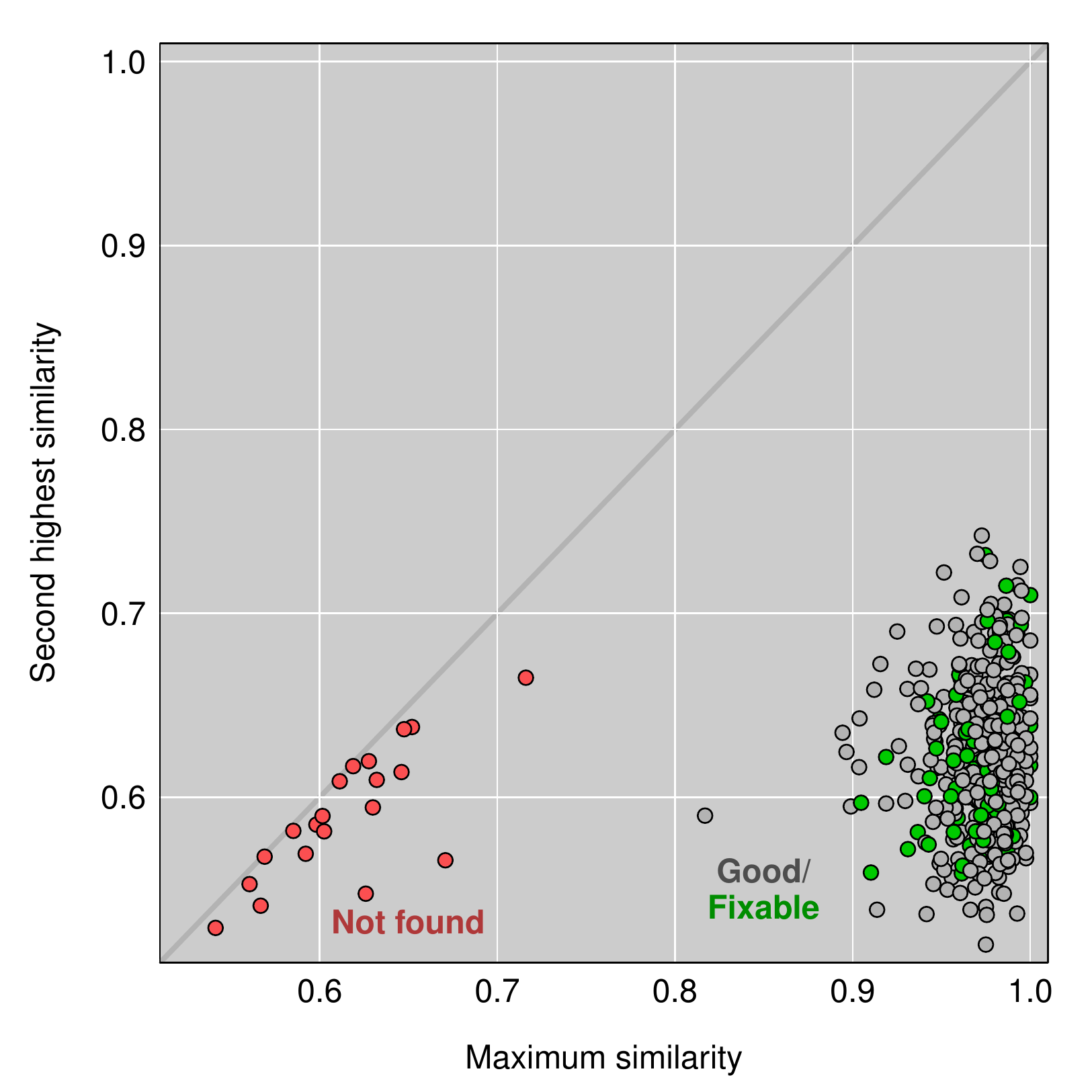}}

\caption{Second highest similarity (proportion matches between
  observed and inferred eQTL genotypes, combined across tissues)
  versus maximum similarity for the DNA samples.  The diagonal gray
  line corresponds to equality.  Gray points correspond to DNA samples
  that were correctly labeled.  Green points correspond to sample
  mix-ups that are fixable (the correct label can be determined).
  Red points comprise both samples mix-ups that cannot be corrected
  as well as samples that may be correct but cannot be verified as no
  expression data is available.
}
\end{figure}

\clearpage

\begin{figure}[p]
\centerline{\includegraphics[width=0.85\textwidth]{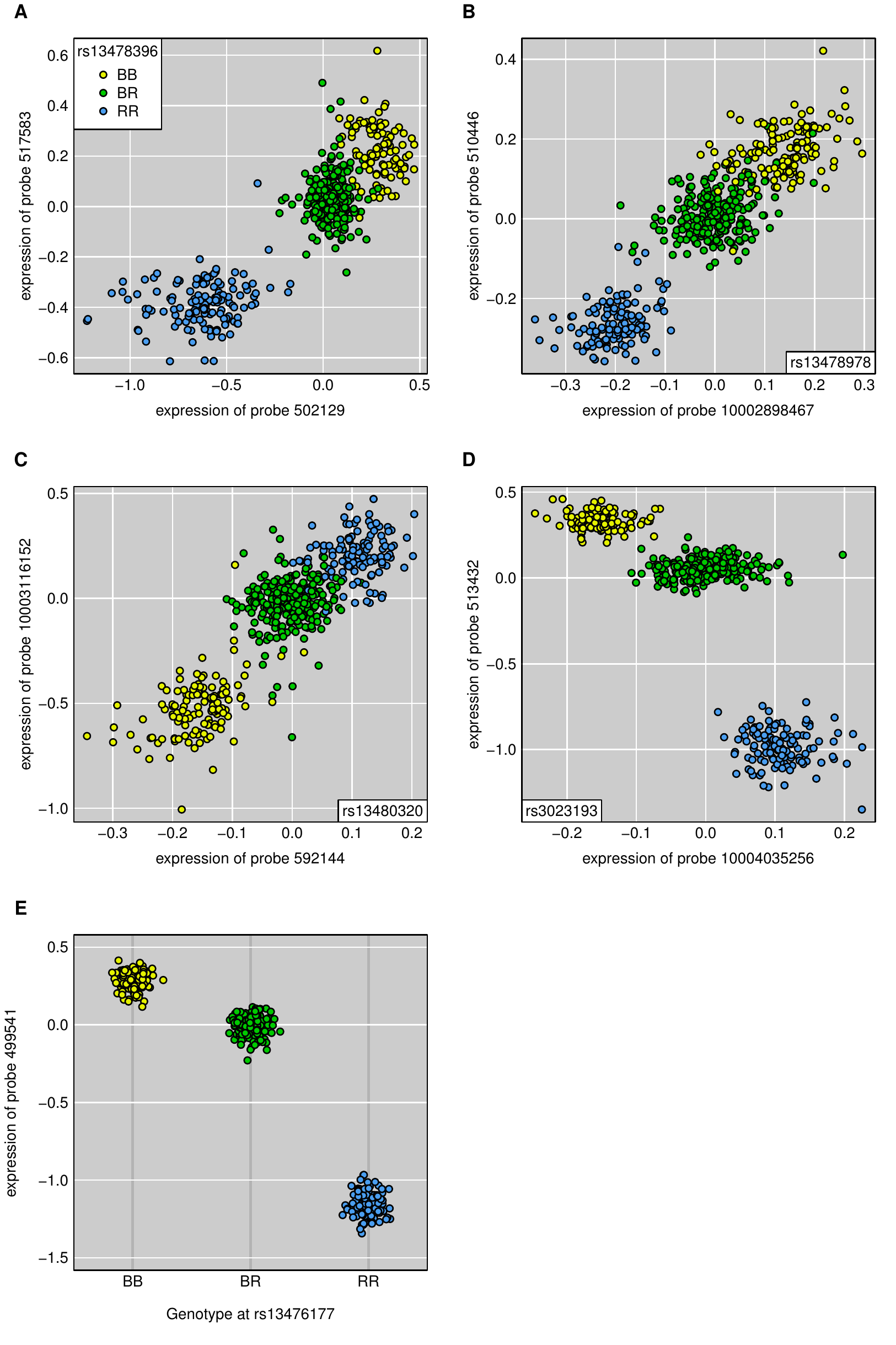}}

\caption{Panels A-D contain the example scatterplots of islet expression for
  pairs of probes at the same genomic location, as in Figure~S13,
  following correction of the sample mix-ups. Panel E contains the plot
  of islet expression vs observed genotype for an example probe, as in
  Figure~5, following correction of the sample mix-ups.
}
\end{figure}

\clearpage

\begin{figure}[p]
\centerline{\includegraphics{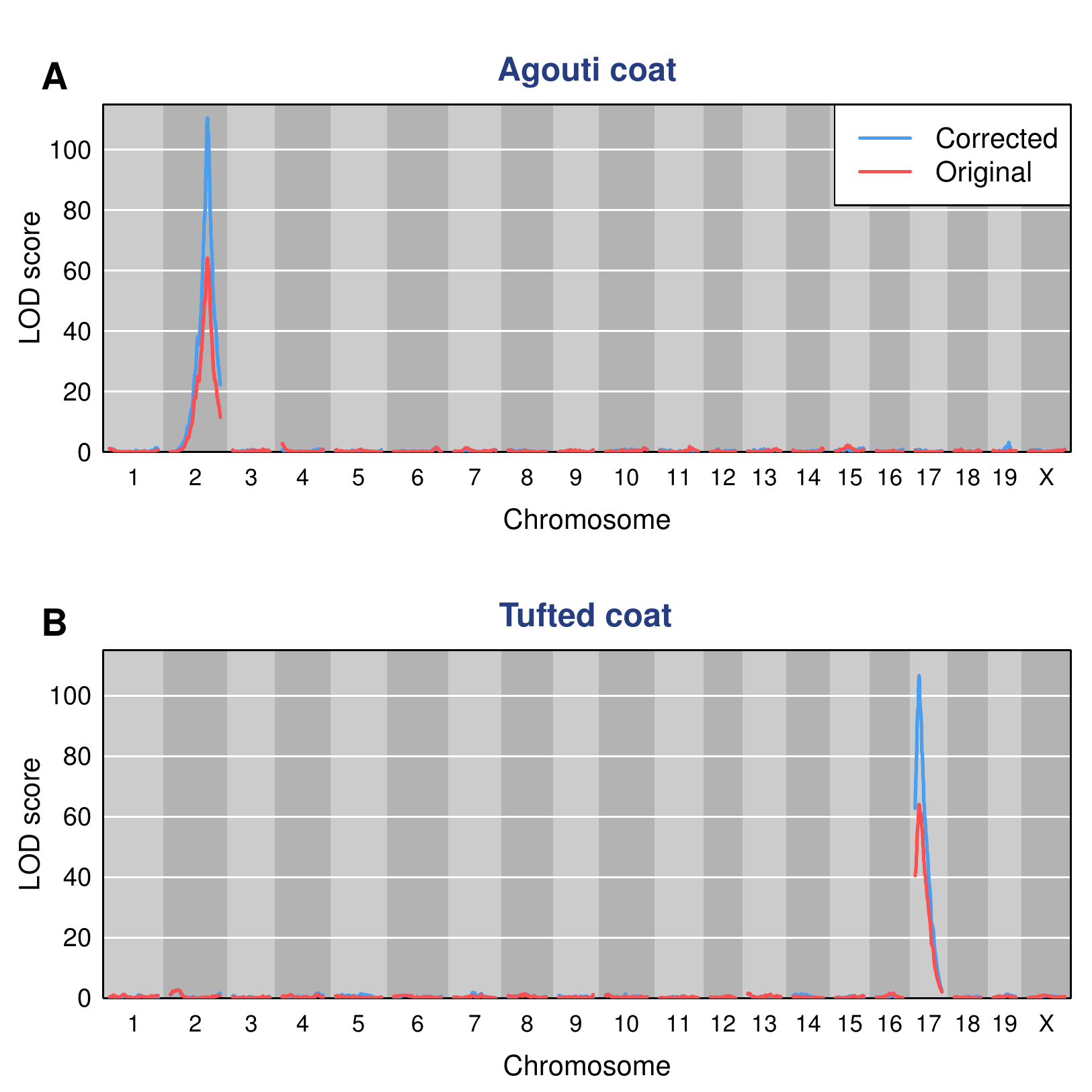}}

\caption{LOD curves for agouti (A) and tufted (B) coat traits with the
  original data (red) and after correction of the sample mix-ups
  (blue).
}
\end{figure}

\clearpage

\begin{figure}[p]
\centerline{\includegraphics{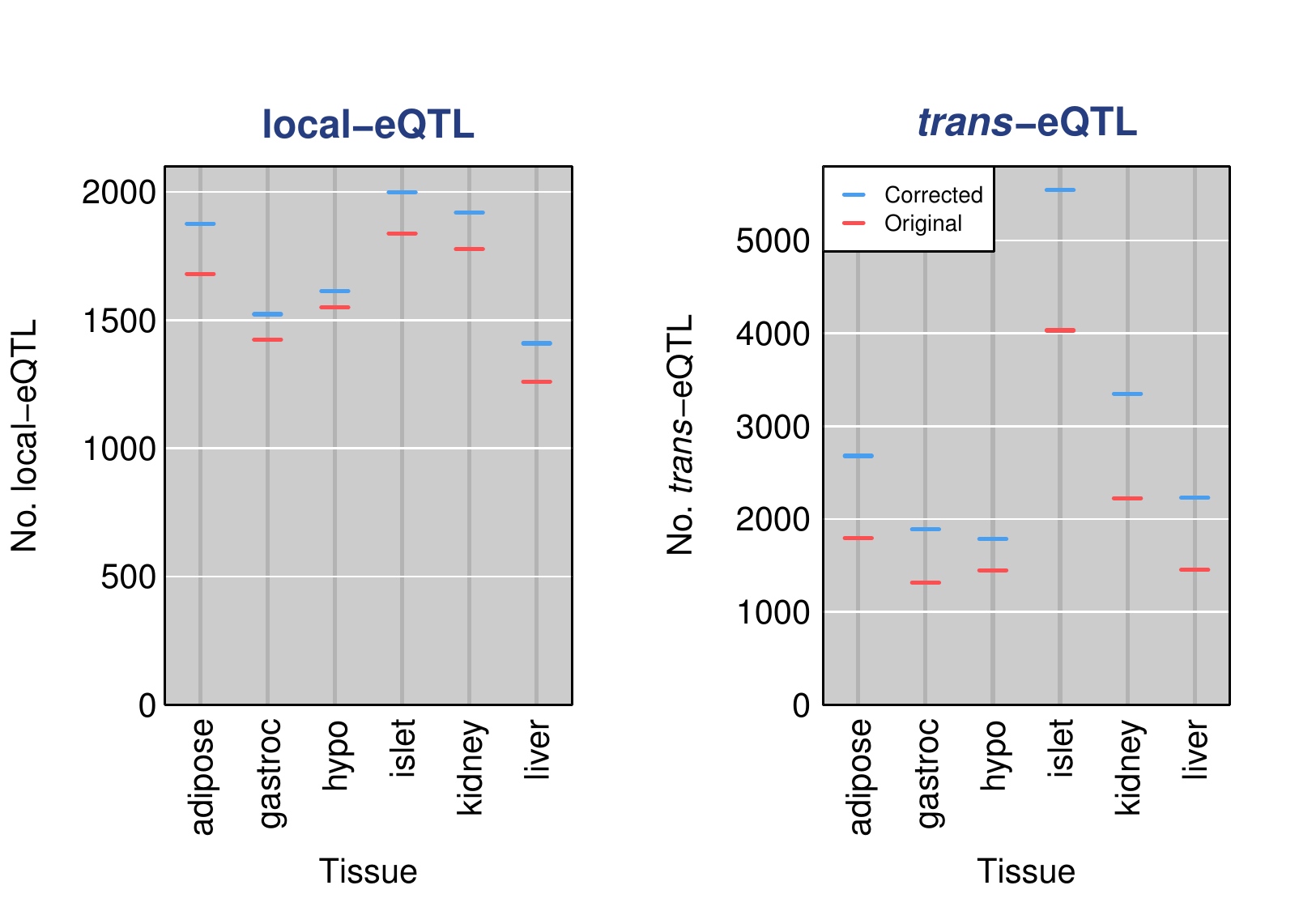}}

\caption{  Numbers of identified local- and \emph{trans}-eQTL with LOD $\ge$ 10,
  with the original data (red) and after correction of the sample
  mix-ups (blue), across 37,797 array probes with known genomic
  location.  An eQTL was considered local if the 2-LOD support
  interval contained the corresponding probe; otherwise it was
  considered \emph{trans}.
}
\end{figure}

\clearpage

\renewcommand{\arraystretch}{1.5}
\captionsetup{font=normalsize}

\begin{table}[p]
\centering
\caption{Duplicate DNA samples}
\begin{tabular}{ccccc}
  \hline
Mouse 1 & Mouse 2 & No. matches & No. typed markers & \% mismatches \\
  \hline
Mouse3259 & Mouse3269 & 2017 & 2022 & 0.2 \\
  Mouse3267 & Mouse3362 & 1933 & 1966 & 1.7 \\
  Mouse3287 & Mouse3290 & 2012 & 2016 & 0.2 \\
  Mouse3317 & Mouse3318 & 1964 & 1996 & 1.6 \\
  Mouse3353 & Mouse3354 & 2026 & 2031 & 0.2 \\
  Mouse3553 & Mouse3559 & 1998 & 2008 & 0.5 \\
   \hline
\end{tabular}
\end{table}

\clearpage

\begin{table}[p]
\centering
\caption{Numbers of gene expression arrays}
\begin{tabular}{cccc}
  \hline
Tissue & \# arrays & \# omitted & \# kept \\
  \hline
adipose & 497 & 4 & 493 \\
  gastroc & 498 & 2 & 496 \\
  hypo & 494 & 1 & 493 \\
  islet & 499 & 1 & 498 \\
  kidney & 482 & 1 & 481 \\
  liver & 491 & 1 & 490 \\
   \hline
\end{tabular}
\end{table}

\clearpage

\begin{table}[p]
\centering
\caption{Numbers of probes, for each tissue pair, with large between-tissue correlation}
\begin{tabular}{cccccc}
  \hline
Tissue 1 & Tissue 2 & corr $>$ 0.70 & corr $>$ 0.75 & corr $>$ 0.80 & corr $>$ 0.90 \\
  \hline
adipose & gastroc & 199 & 143 & 99 & 30 \\
  adipose & hypo & 110 & 72 & 50 & 7 \\
  adipose & islet & 216 & 159 & 106 & 38 \\
  adipose & kidney & 255 & 186 & 135 & 51 \\
  adipose & liver & 159 & 113 & 79 & 19 \\
  gastroc & hypo & 79 & 55 & 43 & 10 \\
  gastroc & islet & 180 & 132 & 92 & 33 \\
  gastroc & kidney & 219 & 164 & 109 & 43 \\
  gastroc & liver & 149 & 102 & 71 & 23 \\
  hypo & islet & 127 & 82 & 57 & 10 \\
  hypo & kidney & 131 & 92 & 60 & 17 \\
  hypo & liver & 63 & 46 & 33 & 6 \\
  islet & kidney & 269 & 200 & 146 & 42 \\
  islet & liver & 152 & 97 & 64 & 24 \\
  kidney & liver & 245 & 155 & 106 & 30 \\
   \hline
\end{tabular}
\end{table}

\clearpage

\begin{table}[p]
\caption{Genotype versus phenotype at the agouti locus}
\begin{center}
\begin{tabular}{ccccccc} \hline
 &\hspace*{5mm}& \multicolumn{2}{c}{Original} &\hspace*{5mm}& \multicolumn{2}{c}{Corrected} \\ \cline{3-4}\cline{6-7}
 &\hspace*{5mm}& \multicolumn{2}{c}{Coat color} &\hspace*{5mm}& \multicolumn{2}{c}{Coat color} \\ \cline{3-4}\cline{6-7}
Chr 2 genotype && Tan & Black && Tan & Black \\ \hline
 BB && 26 & 114 && 5 & 126 \\
 BR && 249 & 15 && 255 & 2 \\
 RR && 88 & 6 && 92 & 0 \\ \hline
\end{tabular}

\bigskip

B = B6 allele; R = BTBR allele
\end{center}
\end{table}

\clearpage

\begin{table}[p]
\caption{Genotype versus phenotype at the tufted locus}
\begin{center}
\begin{tabular}{ccccccc} \hline
 &\hspace*{5mm}& \multicolumn{2}{c}{Original} &\hspace*{5mm}& \multicolumn{2}{c}{Corrected} \\ \cline{3-4}\cline{6-7}
 &\hspace*{5mm}& \multicolumn{2}{c}{Tufted coat} &\hspace*{5mm}& \multicolumn{2}{c}{Tufted coat} \\ \cline{3-4}\cline{6-7}
Chr 17 genotype && No & Yes && No & Yes \\ \hline
 BB && 151 & 7 && 153 & 0 \\
 BR && 258 & 9 && 256 & 0 \\
 RR && 21 & 92 && 4 & 106 \\ \hline
\end{tabular}

\bigskip

B = B6 allele; R = BTBR allele
\end{center}

\end{table}

\end{document}